\documentclass[aip,jcp,amsmath,amssymb,reprint]{revtex4-2}
\usepackage{color}
\usepackage[dvipsnames,svgnames,table]{xcolor}
\usepackage[colorlinks=true,linkcolor=blue,urlcolor=black,citecolor=blue]{hyperref}
\usepackage{graphicx}
\usepackage{bmpsize}
\usepackage{dcolumn}
\usepackage{bm}
\usepackage{xcolor}
\usepackage{enumerate}
\usepackage{booktabs}
\usepackage{threeparttable}
\usepackage{amsmath}
\usepackage[utf8]{inputenc}
\usepackage[T1]{fontenc}
\usepackage{mathptmx}
\usepackage{etoolbox}
\usepackage{natbib}
\usepackage[normalem]{ulem}

\makeatletter
\def\@email#1#2{
 \endgroup
 \patchcmd{\titleblock@produce}
  {\frontmatter@RRAPformat}
  {\frontmatter@RRAPformat{\produce@RRAP{*#1\href{mailto:#2}{#2}}}\frontmatter@RRAPformat}
  {}{}
}
\makeatother

\newcommand{\qed}{\hfill {\rule{0.7em}{0.7em}}}

\newtheorem{lemma}{Lemma}
\newtheorem{theorem}{Theorem}
\DeclareMathOperator*{\argmin}{argmin}

\begin{document}
\title{Information encoding in spherical DFT}
\author{Sol Samuels}
\affiliation{Departments of Physics \& Astronomy and Mathematics \& Statistics, University of New Mexico, Albuquerque, NM USA}
\author{Chance M. Baxter}
\affiliation{Nanoscience and Microsystems Engineering Program, University of New Mexico, Albuquerque, NM USA}
\author{Susan R. Atlas}
\email{susier@unm.edu} 
\affiliation{Departments of Chemistry \& Chemical Biology and Physics \& Astronomy, and Center for Quantum Information and Control, University of New Mexico, Albuquerque, NM USA}

\date{\today}

\begin{abstract}

Spherical density functional theory (DFT) is a reformulation of the classic theorems of DFT, in which the role of the total density of a many-electron system is replaced by a \textit{set} of sphericalized densities, constructed by spherically-averaging the total electron density about each atomic nucleus.  In Hohenberg-Kohn DFT and its constrained-search generalization, the electron density suffices to reconstruct the spatial locations and atomic numbers of the constituent atoms, and thus the external potential.  However, the original proofs of spherical DFT require knowledge of the atomic locations at which each sphericalized density originates, in addition to the set of sphericalized densities themselves. In the present work, we utilize formal results from geometric algebra---in particular, the subfield of distance geometry---to show that for Coulombic systems this spatial information is encoded within the ensemble of sphericalized densities themselves, and does not require independent specification. Consequently, the set of sphericalized densities uniquely determines the total external potential of the system, exactly as in Hohenberg-Kohn DFT.  This theoretical result is illustrated through numerical examples for LiF and for glycine, the simplest amino acid.  In addition to establishing a sound practical foundation for spherical DFT as applied to Coulombic systems, the extended theorem provides a rationale for the use of sphericalized atomic basis densities---rather than orientation-dependent basis functions---when designing classical or machine-learned potentials for atomistic simulation.
\end{abstract}

\maketitle

\section{\label{sec:Introduction} Introduction}

Density functional theory (DFT), established more than half a century ago through the theorems of Hohenberg and Kohn (HK) \cite{hohenberg1964} and Kohn and Sham (KS),\cite{kohn1965} provides the theoretical and practical foundation for the majority of contemporary molecular and condensed matter electronic structure calculations.  
In addition to yielding remarkably accurate quantum mechanical results for ground and excited state properties at a fraction of the computational cost of wavefunction-based quantum chemistry techniques, DFT is now routinely employed in generating large, high-throughput quantum mechanical training datasets for developing machine learning models of classical dynamical potentials.\cite{smith2017, takamoto2022b, chmiela2023accurate, khan2024quantummechanicaldataset836k}  The ability to express the total energy of a quantum system---including spatial and statistical (Pauli) electron correlation---solely as a functional of its total density $\rho({\bf r})$, also provides a powerful mechanism for dynamically coupling the electronic and atomistic length scales.\cite{valone2006Electron,bartok2013representing, ghasemi2015interatomic, brockherde2017bypassing, zhang2019embedded, Atlas2021, behler2021four, musil2021physics} 

Against this backdrop, Theophilou \cite{theophilou2018} proposed a surprising reformulation of ground state density functional theory, in terms of a \textit{set of localized, spherically-averaged densities} $\{\bar{\rho}_i(r_i; {\bf R}_i)\}, i=1,\ldots M$, centered about the $M$ respective nuclei of a many-electron atomic system.  The $\bar{\rho}_i(r_i; {\bf R}_i)$ were defined as\cite{fn3}
\begin{equation}
    \bar{\rho}_i(r_i; {\bf R}_i) \equiv \int \rho({\bf r}_i)\, d\Omega_i,
    \label{eq:spherical-den}
\end{equation}
where ${\bf r}_i = {\bf r} - {\bf R}_i$, $r_i$ is the radial distance from nucleus $i$ in the local coordinate system associated with atom $i$ located at ${\bf R}_i$, and $\Omega_i$ is the angular component of the differential volume in spherical polar coordinates centered about atom $i$.  The set of \textit{localized} 
$\{\bar{\rho}_i(r_i; {\bf R}_i)\}, i=1,\ldots M$ was shown to be an effective 
replacement for the single, \textit{global} electron density $\rho({\bf r})$ of standard DFT, as a direct consequence of the spherical symmetry of the external potential $v_i({\bf r};{\bf R}_i)$ associated with the $i$th atom centered at ${\bf R}_i$:
\begin{equation}
    v_i({\bf r};{\bf R}_i) = - \frac{Z_i}{|{\bf r} - {\bf R}_i|}.
    \label{eq:local-v}
\end{equation}

Shortly after Theophilou's demonstration of spherical DFT for Coulombic systems, Nagy proposed an alternative proof \cite{nagy2018} motivated by E.~Bright Wilson's famous observation \cite{Handy1996} that DFT could be understood intuitively in terms of Kato's theorem.\cite{Kato1957,Steiner1963} Wilson noted that the nuclear cusps in the total electron density $\rho({\bf r})$ serve to identify the set of nuclear locations $\{{\bf R}_i\}$ in the molecule, while the derivatives of the electron density at each cusp location yield the corresponding nuclear charges:
\begin{equation}
    Z_i = - \frac{1}{2\rho({\bf r} = {\bf R}_i)} \left. \frac{\partial \rho({\bf r})}
    {\partial r}\right |_{{\bf r}={\bf R}_i}.
    \label{eq:Kato}
\end{equation}
Together, this information suffices to reconstruct the total external potential $v({\bf r}) = \sum_{i=1}^M v_i({\bf r};{\bf R}_i)$, so that   $\{\bar{\rho}_i(r_i)\}, i=1,\ldots M \Rightarrow v({\bf r})$, the spherical DFT analog of the familiar shorthand statement of the Hohenberg-Kohn theorem, $\ \rho({\bf r}) \Rightarrow v({\bf r})$. In addition, the constraint requiring conservation of the total number of electrons $N$ in the system \cite{hohenberg1964,nagy2018} is satisfied by the integral of $\rho({\bf r})$ over all space.  Nagy's proof of spherical DFT utilized an analog of this argument applied to the individual sphericalized densities. In the same work, Nagy proposed a generalization of Theophilou's theorem for non-nuclear external potentials taking the form of a sum of terms, where each term depends only on the distance from a nucleus (not necessarily Coulombic in form.) The generalization was based on Levy's constrained search formulation of DFT.\cite{levy1979} The theory was subsequently extended to individual excited states \cite{nagy2024density} and ensembles.\cite{nagy2025ensemble}

The notion that a set of sphericalized densities can serve as a rigorous replacement for standard total-density DFT seems at once remarkable and counterintuitive.
However, it must be emphasized that both proofs rely upon a key assumption, namely, that each spherically-averaged nuclear density is ``tagged'' with knowledge of the nuclear location at which it was computed. In Theophilou's proof, this makes it possible to establish a 1-1 correspondence between the local external potential and local sphericalized density along the lines of the original Hohenberg-Kohn proof; in Nagy's proof, Kato's theorem is used to deduce the nuclear charge corresponding to the known nuclear location of a given sphericalized density, allowing a unique reconstruction of the local external potential.  By definition, each sphericalized density separately satisfies the constraint of reproducing the total number of electrons $N$ in the system,\cite{hohenberg1964,nagy2018} since $\int r_i^2\, \bar{\rho}_i(r_i; {\bf R}_i)\, dr_i = N\ \forall\ i = 1,\ldots M$.

In the present work we show that for atomic systems with Coulomb potentials there exists additional information about the electronic structure of the system already encoded within the set of spherical densities, so that {\it a priori} knowledge of the originating nuclear location tags is not required in order to prove that $\{\bar{\rho}_i(r_i)\}, i=1,\ldots M \ \Rightarrow\ v({\bf r}).$  
We thus establish spherical DFT more generally as fully equivalent to ground state HK DFT for Coulombic systems. The proof utilizes formal results from distance geometry,\cite{crippen1978stable, havel1983theory, dokmanic2015euclidean} a subdomain of geometric algebra \cite{havel1998distance,liberti2014euclidean} that plays an essential role in protein structure determination from NMR measurements of intramolecular distances,\cite{wuthrich2003nmr, havel1985evaluation} as well as applications to other macromolecular and nanoscale 3D structure problems.\cite{lazar2020distance, billinge2018recent} 

The organization of the paper is as follows. In Section~\ref{sec:Statement} we state the ground state spherical DFT version of the Hohenberg-Kohn theorem as formulated by Theophilou, and outline his proof, as well as the alternative proofs by Nagy. This provides the necessary background for discussing the extended proof presented here.  In Section~\ref{sec:Encoding} we show how the additional information in the form of nuclear cusps encoded within the atom-centered spherical DFT densities enables a generalized proof using distance geometry methods. The practical application of the extended spherical DFT theorem using Euclidean distance matrix analysis\cite{dokmanic2015euclidean} is illustrated in Section~\ref{sec:Num} for the heteronuclear diatomic LiF and the amino acid glycine, with numerical examples for the cases where either exact or approximate distance matrix information is available. The paper concludes in Section~\ref{sec:Discussion} with a discussion of the relationship of spherical DFT to the atom-in-molecule problem,\cite{theophilou2018,nagy2019} and implications for the design of atomistic potentials based on spherically-symmetric atomic basis densities\cite{Atlas2021} and machine learning techniques.

\section{Statement and proofs of spherical DFT}
\label{sec:Statement}
As background for the extension proposed in Section~\ref{sec:Encoding}, we review the proofs establishing the ground state spherical DFT analog of the Hohenberg-Kohn theorem.  Since it will be key to the arguments presented below, we use notation that preserves the parametric dependence on nuclear locations $\{{\bf R}_i\}$ for both the sphericalized densities and corresponding nuclear potentials. The original theorem is then stated as:\cite{theophilou2018}
\begin{quote}
    The set of spherical densities $\{\bar{\rho}_i(r_i; {\bf R}_i)\}$, $i= 1,\ldots M$ uniquely determines the external potential $v({\bf r})$ of the system.
\end{quote}  
In the present work, we focus exclusively on the Hohenberg-Kohn formulation of spherical DFT.  The derivations of the  corresponding Kohn-Sham and Euler equations can be found in Refs.~\onlinecite{theophilou2018,nagy2018,nagy2019,nagy2024density}.

\subsection{Theophilou's proof}
The original statement and proof of spherical DFT is due to Theophilou.\cite{theophilou2018}  The proof is based on a {\it reductio ad absurdum} argument similar to that of Hohenberg and Kohn (HK),\cite{hohenberg1964} but with global quantities replaced by local ones:
\begin{equation}
    \rho({\bf r}) \rightarrow \bar{\rho}_i(r_i; {\bf R}_i);\ \ \ \ v({\bf r}) \rightarrow v_i({\bf r};{\bf R}_i).
\end{equation}
Note that the local densities and potentials are each parameterized by their respective nuclear coordinates ${\bf R}_i$.

Consider a given atom $i$. The potential of Eq.~(\ref{eq:local-v}) is spherically symmetric about the nucleus located at ${\bf R}_i$. Expanding $\rho({\bf r})$ in the complete set of spherical harmonics $Y_{lm}(\Omega_i)$ centered about ${\bf R}_i$ gives:
\begin{equation}
    \rho({\bf r}) = \rho_0^i (r_i;{\bf R}_i) + \sum_{l>0; m} \rho_{lm}(r_i) Y_{lm}(\Omega_i),
\end{equation}
where the $\rho_{lm}$ are expansion coefficients.  
Spherically averaging over all angles $d\Omega_i$, the only contribution that remains is the spherical component of the density, $\rho_0^i(r_i;{\bf R}_i)$, since for $l > 0$, $\int Y_{lm}(\Omega_i)\, d\Omega_i = 0$. Clearly, $\rho_0^i(r;{\bf R}_i) = \bar{\rho}_i(r_i; {\bf R}_i)$ as defined in Eq.~(\ref{eq:spherical-den}) above, and as a consequence, it can be shown that (see Appendix~B of Ref.~[\onlinecite{theophilou2018}]):
\begin{equation}
\int v_i({\bf r};{\bf R}_i) \rho({\bf r})\, d{\bf r} = \int_0^\infty r_i^2 \left ( -\frac{Z_i}{r_i} \right )\, 
\bar{\rho}_i(r_i; {\bf R}_i)\, d{r_i}.
\label{eq:AppB}
\end{equation}

The original HK proof by contradiction assumes that there exist two distinct external potentials $v({\bf r})$ and $v'({\bf r})$, with corresponding distinct ground state wavefunctions $\Psi$ and $\Psi'$ that give rise to the identical density $\rho({\bf r})$.  By application of the standard wavefunction-based variational principle, 
HK show that the statements $E < E'$ and $E' < E$ must both be true, a contradiction.  Thus, $\rho({\bf r})$ uniquely determines $v({\bf r})$.

In spherical DFT, the contradiction is established on a term-by-term basis, using the result quoted in Eq.~(\ref{eq:AppB}) to derive the relation\cite{theophilou2018} 
\begin{align}
    \sum_i \int dr_i\, r_i^2\, [v_i'(r_i;{\bf R}_i) &- v_i(r_i;{\bf R}_i)]\times \nonumber \\ 
    [\bar{\rho}_{i,\Psi'}(r_i;{\bf R}_i) &- \bar{\rho}_{i,\Psi}(r_i;{\bf R}_i)] < 0. 
    \label{eq:contra}
\end{align}
This is done by applying the variational principle to the total external potential $v({\bf r})$ with ground state $\Psi$, and total external potential $v'({\bf r})$ with ground state $\Psi'$, just as in HK, but now with $M$ successive definitions of $v'({\bf r}) = \sum_{i\ne j} v_i({\bf r};{\bf R}_i) + v'_j({\bf r};{\bf R}_j)$.  Since each sphericalized density $\bar{\rho}_{i,\Psi}(r_i;{\bf R}_i)$ = $\int \rho_{\Psi}({\bf r};{\bf R}_i)\, d\Omega_i$, and derives from the same total electron density $\rho_\Psi({\bf r})$ associated with total wavefunction $\Psi$, if the total densities associated with $\Psi$ and $\Psi'$ are assumed to be identical as in the HK proof, then $[\bar{\rho}_{i,\Psi'}(r_i;{\bf R}_i) - \bar{\rho}_{i,\Psi}(r_i;{\bf R}_i)]$ must equal 0 for all $i$, contradicting the inequality in Eq.~(\ref{eq:contra}).  This implies that on a term-by-term basis, the sphericalized densities uniquely determine the corresponding spherically-symmetric potential components $v_i(r_i;{\bf R}_i)$, and thus their sum, the total external potential $v({\bf r})$:
\begin{equation}
v({\bf r}) = \sum_{i=1}^M v_i({\bf r};{\bf R}_i).
\label{eq:v}
\end{equation} 
The usual HK DFT statement and proof that $\rho({\bf r}) \Rightarrow\, v({\bf r})$ has been converted into the set of spherical DFT statements $\bar{\rho}_i(r_i; {\bf R}_i) \Rightarrow\, v_i(r_i;{\bf R}_i)$, $i= 1,\ldots M$, collectively implying that $\{\bar{\rho}_i(r_i; {\bf R}_i)\} \Rightarrow\, v({\bf r})$. 

\subsection{Nagy's approach to spherical DFT}
Nagy proposed two alternative formulations of spherical DFT. The first utilizes Kato's theorem\cite{Kato1957} as extended to electron densities by Steiner.\cite{Steiner1963} Like Theophilou's proof, Nagy's approach is based on a set of atom-by-atom correspondences between spherical densities and spherically-symmetric potentials.   Instead of using Kato's theorem in its original ``global'' form, however, Nagy notes that since the density at an atomic nucleus is equal to its spherically-averaged value there, \textit{i.e.}, 
$\rho({\bf r} = {\bf R}_i) = \bar{\rho}_i(r_i =0; {\bf R}_i)$, Kato's theorem can rewritten as:
\begin{equation}
\left.\frac{\partial \bar{\rho}_i(r_i;{\bf R}_i)} {\partial r_i} \right \rvert_{r_i = 0} = -2Z_i\bar{\rho}_i(r_i = 0).
\end{equation}
Each $Z_i$ is then easily deduced from its corresponding spherically-averaged density $\bar{\rho}(r_i;{\bf R}_i)$.  Since the
$\{{\bf R}_i\}$ are assumed given, this allows the immediate reconstruction of the external potential $v({\bf r})$ of Eq.~(\ref{eq:v}).  Nagy further notes that separately integrating each spherical density $\bar{\rho}_i(r_i;{\bf R}_i)$ about its respective nuclear center must, by definition, yield the total number of electrons $N$:
\begin{equation}
    \int r_i^2\, \bar{\rho}_i(r_i;{\bf R}_i)\,  dr_i = N.
\end{equation}
This suggests that the initial formulation of spherical DFT may contain redundant information.  Indeed, it is this redundancy that enables the extension proposed in Section~\ref{sec:Encoding}.

Nagy's second proposed reformulation of spherical DFT is based on constrained search theory.\cite{levy1979} The original Hohenberg-Kohn universal functional $F_{\rm HK}[\rho]$ is defined as
\begin{equation}
    F_{\rm HK}[\rho] = \langle \Psi [\rho] |\hat{T} + \hat{W} | \Psi [\rho] \rangle,
\end{equation}
where ${\hat T}$ and ${\hat W}$ are the kinetic energy and electron-electron repulsion operators.  The ground state energy $E_0$ is obtained through the minimization
\begin{equation} 
E_0 = \min_\rho E_v[\rho] = \min_\rho \left [ F_{\rm HK}[\rho] + \int v({\bf r}) \rho({\bf r}) d{\bf r} \right ],
\label{eq:HK}
\end{equation}
where $v({\bf r})$ is the external potential. To avoid the so-called $v$-representability problem (potentially searching over densities that include pathological $\rho({\bf r})$ not derivable from any physically-realizable external potential $v({\bf r})$), Levy proposed replacing $F_{\rm HK}[\rho]$ in Eq.~(\ref{eq:HK}) with the alternative functional $Q[\rho]$:
\begin{equation}
   Q[\rho] = \min_{\Psi \rightarrow \rho} \langle \Psi |\hat{T} + \hat{W} | \Psi \rangle.
\end{equation}
This avoids the $v$-representability problem through the nested minimization 
\begin{align}
    E_0 &= \min_\rho \left [ Q[\rho] + \int v({\bf r}) \rho({\bf r}) d{\bf r} \right ] \nonumber \\
              &= \min_\rho \left [ \min_{\Psi \rightarrow \rho} \langle \Psi |\hat{T} + \hat{W} | \Psi \rangle  + \int v({\bf r})\,   \rho({\bf r})\, d{\bf r} \right ]. 
\end{align}
The weaker requirement placed on $\rho({\bf r})$ is that of $N$-representability: the total density $\rho({\bf r})$ must be derivable from \textit{some} total antisymmetric wavefunction $\Psi$. 

Nagy's constrained search formulation of spherical DFT is based on the definition of the functional
\begin{equation}
    Q[\{\bar{\rho}_i(r_i;{\bf R}_i)\}] = \min_{\Psi \rightarrow \{\bar{\rho}_i(r_i;{\bf R}_i)\}} \langle \Psi |\hat{T} + \hat{W} | \Psi \rangle ,
\label{eq:NQ}
\end{equation}
where the minimization in Eq.~(\ref{eq:NQ}) is over all $N$-electron antisymmetric wavefunctions $\Psi$ yielding a specified set of sphericalized densities $\{\bar{\rho}_i(r_i;{\bf R}_i)\}$, $i=1,...M$, each separately integrating to $N$.  (In what follows, when using the $\{\ \}$ notation to denote the collection of sphericalized densities, the complete set $i=1,...M$ is understood.)  As noted above, the constrained search formulation allows a generalization of the original proofs to non-Coulombic systems.
Assuming that such $\Psi$ can be found, Eq.~(\ref{eq:NQ}) implies the corollary
$\{\bar{\rho}_i(r_i;{\bf R}_i)\} \Rightarrow \rho({\bf r})$. \cite{nagy2018,nagy2019} 
Moreover, due to the spherically-symmetric form of the potential at each nucleus, the total ground state energy of the system can be computed by minimizing the energy with respect to sets of sphericalized densities:\cite{nagy2018}
\begin{align}
    E_0 &= \min_{\{\bar{\rho}_i(r_i;{\bf R}_i)\}} \biggl [ Q[\{\bar{\rho}_i(r_i;{\bf R}_i)\}]  \nonumber \\ 
        &+ \sum_{i=1}^M \int r_i^2\, \bar{\rho}_i(r_i;{\bf R}_i)\, v_i(r_i;{\bf R}_i)\, dr_i \biggr ].
\end{align}

\section{Extended spherical DFT 
\label{sec:Encoding}}  
In this section we show that there exists sufficient information in the form of nuclear cusps already encoded within the set of $M$ sphericalized densities 
$\{\bar{\rho}_i(r_i)\}$, so that the total external potential can be reconstructed without requiring {\it a priori} knowledge of the original locations $\{{\bf R}_i\}$ at which the spherical densities were computed.  That is, we show that $\{\bar{\rho}_i(r_i)\}$ $\Rightarrow v({\bf r})$ directly, with the parametric dependence on nuclear position omitted. This is accomplished via the determination of atom-to-atom distances from Kato peak signatures within each sphericalized density distribution. The complete set of atom-to-atom distances can be
utilized to uniquely determine the nuclear coordinates. We thus generalize the original results of Theophilou and Nagy for Coulombic systems.

The proof proceeds through a series of results established over the following subsections. Before doing so, however, two remarks are in order.

The first relates to non-nuclear maxima (NNM) in the electron density (also known as \textit{non-nuclear attractors} (NNA)). Non-nuclear maxima have long been known to arise in various molecular systems, notably ${\rm Li}_2$\cite{dunning2024electronic,anderson2021non} and other homomuclear diatomics.\cite{pendas1999non}  In a recent work,\cite{anderson2021non} Anderson {\it et al.}~considered the right-hand-side of Eq.~(\ref{eq:Kato}), which they refer to as the {\it Kato-ratio}, and showed that where the electron density is smooth and differentiable, the Kato-ratio is identically zero.  Consequently, Nagy's Kato cusp condition proof holds even in cases where the electron density exhibits non-nuclear maxima, and only true nuclear cusps need to be considered in establishing spherical DFT.  Early work by Pack and Byers-Brown on molecular cusps is consistent with this conclusion.\cite{pack1966cusp}

The second point concerns numerical effects. Kato's theorem (Eq.~(\ref{eq:Kato})) implies that the electron density close to the nucleus, and its sphericalization, can be described by an exponential cusp at $r=0$, expressed as
\begin{equation}
    \bar{\rho}_i(r)= \bar{\rho}_i(0)e^{-2Z_i r}\hspace*{.05in}{\rm as\ } r \rightarrow 0.
    \label{eq:Kato2}
\end{equation}
This expression was tested in atoms by Liu {\it et al.},\cite{liu1995} with further studies of cusp topology for excited and charge states performed by Nagy and co-workers.\cite{nagy2020,nagy2020b,nagy2001}  It was subsequently used as a constraint in modeling radially-symmetric neutral and ionic electron densities for implementing the ensemble charge-transfer embedded-atom method.\cite{amokwao2020, Atlas2021} However, electron densities computed using Gaussian basis sets, and their spherical averages, can smooth out the strict exponential behavior of the nuclear cusps, so that they are no longer discernible as distinct peaks, or are masked under a single, broadened peak.  These numerical effects are observed here for molecular hydrogens as previously described by others,\cite{anderson2021non} for atoms (C, N) that happen to lie in close proximity to one another, and in the Gaussian shapes of the detected peaks.  Nevertheless, the exact analytic results for the various cusps, as derived in the following section, still hold.  As shown in Section~III.B, the ability to identify the complete set of $M(M-1)/2$ pairwise distances between all nuclei in a molecule implies that the 3D coordinate locations of all nuclei can be reconstructed.  When combined with Theophilou's or Nagy's original arguments, this suffices to prove the extended spherical DFT analog of the Hohenberg-Kohn theorem for Coulombic systems.

\subsection{Spherical averaging preserves nuclear cusps}
In this section we show that spherical averaging of the total electron density $\rho({\bf r})$ about any atomic center ${\bf R}_i$ in an $M$-center molecule preserves the cusp behavior associated with all remaining $M-1$ atomic centers. 

\begin{lemma}
  Spherical averaging of the total electron density $\rho({\bf r})$ of a two-center molecule about one of its atomic centers preserves the cusp behavior at both centers of the molecule.
\end{lemma}

\begin{figure}
    \centering
    \includegraphics[width=0.5\linewidth]{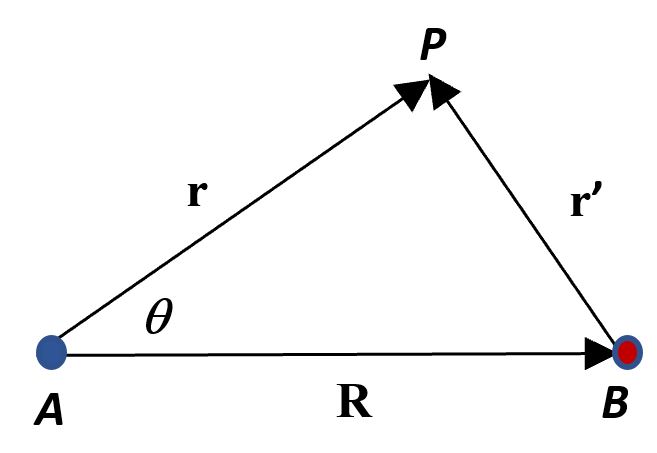}
    \caption{Spherical averaging geometry, illustrating the relation between 3D vectors ${\bf r}$, ${\bf R}$, and ${\bf r'}$.}
    \label{fig:geom}
\end{figure}

\noindent
{\bf Proof.}  Consider the two-center molecule AB, with atom B located a distance $R$ from atom A.  Without loss of generality, we fix A to be located at the origin (0,\, 0,\, 0) and B at (0,\, 0,\, $R$) along the $z$ axis.  The geometry is illustrated in Fig.~\ref{fig:geom}, where the coordinate to be spherically averaged over is the vector ${\bf r}$, and we use the spherical polar coordinate definition
\begin{align}
    x &= r \cos \theta \cos \phi  \nonumber \\  
    y &= r \sin \theta \sin \phi \nonumber \\ 
    z &= r \cos \theta .
\end{align}

\noindent
The vector ${\bf r}$ corresponds to the location of an arbitrary point \textit{\textbf{P}} whose density will be angle-averaged in computing the sphericalized density
\begin{equation}
    \bar{\rho}_A (r) = \int \rho({\bf r})\, \sin \theta \, d\theta d\phi.
    \label{eq:formal-avg}
\end{equation}

Note that the analogous sphericalized density for atom B is
\begin{equation}
    \bar{\rho}_B (r') = \int \rho({\bf r'})\, \sin \theta' \, d\theta' d\phi',
    \label{eq:formal-avg2}
\end{equation}
and the set of sphericalized densities for this simple diatomic molecule is $\{\bar{\rho}_i\}=\{\bar{\rho}_A,\bar{\rho}_B\}$.

We begin by writing the total electron density as the sum of contributions from A and B expressed in terms of local, sphericalized models of their constituent nuclear cusps, plus a remainder term ${\tilde \rho}({\bf r})$:

\begin{align}
\rho({\bf r}) &= \rho_0^A(r)f_{\rm A}(r) + \rho_0^B(|{\bf r} - {\bf R}|)f_{\rm B}(|{\bf r} - {\bf R}|) + {\tilde \rho}({\bf r}) \nonumber \\
&= \rho_0^A(r)f_{\rm A}(r) + \rho_0^B(r') f_{\rm B}(r') + {\tilde \rho}({\bf r}),
\label{eq:model-rho}
\end{align}
where $\mathbf{r}'=\mathbf{r}-\mathbf{R}$. Here $\rho_0^A(r) = {\cal A} e^{-\alpha_{\rm A} r}$ and $\rho_0^B(r') = {\cal B} e^{-\alpha_{\rm B} r'},$ with
$\alpha_A = 2Z_{\rm B}$, ${\cal A} = \rho_{\rm A}^0(0),$ $\alpha_{\rm B} = 2Z_{\rm B},$ and ${\cal B} = \rho_{\rm B}^0(0)$. $f_{\rm A}(r)$ and $f_{\rm B}(r')$ are chosen to be smooth functions equal to 1 at their respective nuclei, decaying smoothly to 0 at short range.  The ranges can be chosen so as to localize each cusp at a short distance, thus avoiding potential conflict between the short-range exponential behavior at each nucleus and the global, long-range exponential decay of the molecular density: 
\begin{equation} 
\rho(r) \sim e^{-2\sqrt{-2\epsilon_I}r}
\label{eq:long-range}
\end{equation} 
as $r\rightarrow \infty$, where $\epsilon_I$ is the least-negative occupied orbital energy.\cite{handy1969,morrell1975,levy1976,ahlrichs1976,hoffmann1977,Tal1978,katriel1980,levy1984}
Since the only cusps in a molecular density are located at the nuclei,\cite{anderson2021non} the remainder term ${\tilde \rho}({\bf r})$ is a smooth, differentiable function of ${\bf r}$ (no cusps).  The decomposition of Eq.~(\ref{eq:model-rho}) is exact.

Substituting the expression for $\rho({\bf r})$ from Eq.~(\ref{eq:model-rho}) into Eq.~(\ref{eq:formal-avg}):
\begin{align}
    \bar{\rho}_A (r)& = \int \rho_0^A(r) f_{\rm A}(r) \sin \theta d\theta\, d\phi + \int \rho_0^B(r')f_{\rm B}(r') \sin \theta d\theta d\phi \nonumber \\ 
    &+ \int {\tilde \rho}({\bf r}) \sin \theta d\theta\, d\phi.
    \label{eq:expl-avg}
\end{align}
Clearly, since $\rho_0^A(r)$ is already a spherical density, and $f_{\rm A}(r)$ is a smooth function of $r$, the  cusp in $\rho_0^A(r)$ remains upon sphericalizing with respect to coordinate ${\bf r}$ about atom A, and the first term in Eq.~(\ref{eq:expl-avg}) is just $4 \pi \rho_0^A(r)f_{\rm A}(r)$.

Now consider the third, remainder term.  By construction, and in light of the arguments regarding NNMs above, this term is smooth and cusp-free.  Consequently, spherical averaging of this contribution cannot introduce new cusps or attenuate the existing ones at atoms A and B.

It therefore remains to consider the second term, which we denote by ${\cal I}_B$.  Evaluating this term requires referencing the local atom B (primed) coordinate system to the (unprimed) sphericalization coordinate system centered at atom A.  This analysis is performed in Appendix~A. The result is:
\begin{align}
    {\cal I}_B &\equiv \int \rho_0^B(r')f_{\rm B}(r') \sin \theta d\theta d\phi \nonumber \\
            &= 4 \pi \rho_0^B(r')f_{\rm B}(r') + \sum_{l>0} \rho_l^B(r')\, f_{\rm B}(r') \times {\rm poly}(r,r',R)|_{r' \ne 0},
            \label{eq:AppA-result}
\end{align}
where ${\rm poly}(r,r',R)|_{r' \ne 0}$ denotes a polynomial in the variables $r$, $r',$ and $R$, with $r' \ne 0.$  
The first term in Eq.~(\ref{eq:AppA-result}) corresponds to a sphericalization about atom A of a function that depends only on the radial coordinate $r'$ referenced to atom B. This preserves the cusp information at $r' = 0$ associated with the spherical density component at atom B.\cite{nagy2018} 
In the second term, spherical averaging of the cusp at atom B with respect to center A introduces polynomial functions of the distances $r$, $r'$, and internuclear separation $R$ multiplying the $l>0$ components of the spherical harmonic expansion of $\rho_l^B(r')$, with $r' \ne 0$.  Clearly, these polynomial terms cannot destroy the exponential cusp at atom B. Thus, the lemma is established. \qed

\begin{lemma}
    The result in Lemma~1 holds for an arbitrary number of nuclei $M$ in a molecule. Spherical averaging over the total molecular density centered at a given atomic location ${\bf R}_i$ preserves cusp behavior at the $i$th atom, as well as at the remaining $M-1$ atomic nuclei located at $\{{\bf R}_j\}$, $j\ne i$.
\end{lemma}

\noindent
{\bf Proof.}  This is a straightforward generalization of Lemma~1, with the total density expressed as the sum of individual cusp models for each of the $M$ centers, plus a remainder term that is smooth and cusp-free.\cite{anderson2021non}  The angular integration over each atomic cusp model can be performed independently, with its local coordinate system rotated into that of Fig.~\ref{fig:geom} as for the two-center problem.  Consequently, all of the cusps survive angular integration, as before. \qed  

\subsection{Distance geometry and spherical DFT} 
As a consequence of Lemma~2, the sphericalized density centered about the $i$th atom in a molecule exhibits peaks (``bumps'') at the set of distances $\{r_j\}$ corresponding to the remaining $M-1$ atoms. Each sphericalized density thus has the appearance of a series of smoothed cusps superimposed upon a smoothly-decaying background.  This behavior is evident in the numerical results shown below for the spherical density centered at one of glycine's hydrogen atoms (Fig.~\ref{fig:atom6}.)  Since each sphericalization is performed over the entire molecular density, with only the center location changing for each atom in the set, the sphericalized densities each exhibit the expected exponential decay at long range, characteristic of the molecule as a whole,\cite{morrell1975,ahlrichs1976,hoffmann1977,katriel1980,Tal1978,levy1976,levy1984} as previously noted by Nagy.\cite{nagy2020} This is illustrated numerically in Appendix~C.

The analytic existence of the nuclear cusp peaks for Coulombic systems\cite{Kato1957,Steiner1963} provides the key to extending the proof of spherical DFT and eliminating the requirement of knowing the 3D origin of each sphericalized density in the molecular set.
The mathematics of distance geometry originated in work from the early 20th century \cite{dokmanic2015euclidean} and was further developed by Crippen,\cite{crippen1977novel,crippen1978stable} Havel,\cite{havel1983theory} and co-workers (see [\onlinecite{havel1983theory}], [\onlinecite{havel1998distance}], and references therein) for applications to molecular structure. This methodology was applied with considerable success to the determination of protein structure from NMR data,  \cite{havel1984distance,havel1985evaluation,williamson1985solution} leading to the Nobel Prize in Chemistry in 2002.\cite{wuthrich2003nmr}

The history and mathematical formulation of distance geometry is reviewed in [\onlinecite{havel1983theory}] and [\onlinecite{liberti2014euclidean}]. The first of the two results required here states that knowledge of the $N(N-1)/2$ distances between $N$ particles suffices to determine all particle coordinates. Following the proof in [\onlinecite{havel1983theory}] (Havel {\it et al.}'s Theorem 1.1), consider $N$ particles embedded in an $n$-dimensional coordinate space, with $n \ge N$; here $n$ = $3N$. Without loss of generality, the number of independent coordinates needed to represent each particle can be reduced to $N-1$ by transforming coordinates so that the particles lie in an $N-1$-dimensional hyperplane. The total number of values required to specify the particle locations is then $N(N-1)$. This number is further reduced by the number of translational degrees of freedom in the hyperplane, $N-1$, and the number of rotational degrees of freedom, ${N-1 \choose 2}$ = $(N-1)(N-2)/2$ (this corresponds to the number of independent rotational planes that can be formed from all possible choices of two axes in the hyperplane.) This gives a total of $N(N-1) - (N-1) -(N-1)(N-2)/2$ = $N(N-1)/2$ values, corresponding precisely to the number of independent pairwise distances between the $N$ particles. 

As a concrete example, consider $N= 3$ particles embedded in 3D coordinate space ($n=3N = 9$).  Since three points determine a plane, the $z$ coordinates of each particle can be set = 0 so that the particles now live in a (hyper)plane of dimension $N-1 = 2$, and the number of values per particle needed to specify the particle coordinates is just $N-1 = 2$---{\it i.e.}, the $x$ and $y$ values for each particle.  For $N=3$ particles, this gives a total of $N(N-1)=6$ values.  The number of translational degrees of freedom in the hyperplane is $N-1$ = 2, and the number of rotational degrees of freedom is ${3 \choose 2} = 1$.  Thus, the total number of coordinates needed to specify the locations of the particles is reduced to 3, corresponding to the number of independent pairwise distances between the three particles.  

In addition to determining the list of reconstructed nuclear centers $\{{\bf R}_i\}$, $i= 1, \ldots M$, it is necessary to establish the correct mapping back to the set of $M$ sphericalized densities $\{\bar{\rho}_i(r_i)\}$ from which the lists of cusp distances originated. Since the origin of each sphericalized density is at a physical atom, this corresponds to the {\it assigned distance geometry problem}. This is a special case of the general {\it unassigned distance geometry problem} (uDGP),\cite{dokmanic2015euclidean,duxbury2016unassigned,billinge2018recent} as the distances are assigned at one end to the atom about which the sphericalization was performed. In the uDGP, all that is available is a list of distances with no knowledge of where those distances originated. The uDGP can be couched as a graph minimization problem in which the mismatch between trial and actual distance lists associated with different possible mappings is minimized.\cite{duxbury2016unassigned}  When the number of coordinate centers is equal to the number of distances, as is the case here, the mapping is bijective, and the inversion is unique.\cite{duxbury2016unassigned} Translated into the language of spherical DFT, we have:

\begin{lemma}
    Knowledge of the $M$ sphericalized densities $\{\rho_i^0(r_i)\}$, $i = 1,\ldots M$ of an $M$-center molecule suffices to determine all unique $M(M-1)/2$ pairwise distances between the atoms in the molecule.  The resulting (symmetric) distance matrix can be inverted to determine the coordinates of all $M$ nuclei in some 3D coordinate system; these nuclear coordinates are uniquely associated with their originating sphericalized densities.  
\end{lemma}

\noindent
{\bf Proof.} According to Lemma~2, each sphericalized density encodes knowledge of the $M-1$ distances to the remaining atoms in the molecule, identified via the locations of their Kato cusps. From the distance geometry results quoted above for $N=M$ particles, the total of $M(M-1)/2$ independent distances provides the information needed to uniquely specify the coordinates of all atoms and associated sphericalized densities in the $3N$-dimensional embedding coordinate space. \qed 

\vspace*{.1in}
\noindent
In practice, as in the case of experimental NMR data, uncertainties in distance measurements and missing values in the distance matrix can limit the accuracy of coordinate determination, although computational algorithms are available to mitigate these issues, for example through the application of geometric constraints.\cite{havel1983theory,havel1998distance,dokmanic2015euclidean,duxbury2016unassigned,billinge2018recent} 
For purposes of establishing the extended spherical DFT theorem, however, the distances are assumed to be known exactly.

\vspace*{.05in}
\noindent
The main result of the present work can now be stated as:
\begin{theorem}
    {\rm Extended spherical DFT.} The set of sphericalized densities $\{\bar{\rho}_i(r_i)\}, i=1,\ldots M$
    suffices to reconstruct the external potential $v({\bf r})$ of a system of $M$ atoms with  total number of electrons $N$, thus establishing the analog of the Hohenberg-Kohn theorem for spherical DFT in Coulombic systems. {\rm A priori} knowledge of the nuclear center locations associated with each spherical density is not required.
\end{theorem}

\noindent
{\bf Proof.} From Lemma~3, the sphericalized densities encode pairwise distance information that allows the reconstruction of all $M$ nuclear coordinate locations $\{{\bf R}_i\}$, $i= 1, \ldots M$, and their unique assignment to the originating sphericalized densities from which the pairwise distances were identified. The theorem then follows immediately from the previous proofs of Theophilou\cite{theophilou2018} and Nagy.\cite{nagy2018} \qed

\begin{figure*}
\includegraphics[width=1\linewidth]{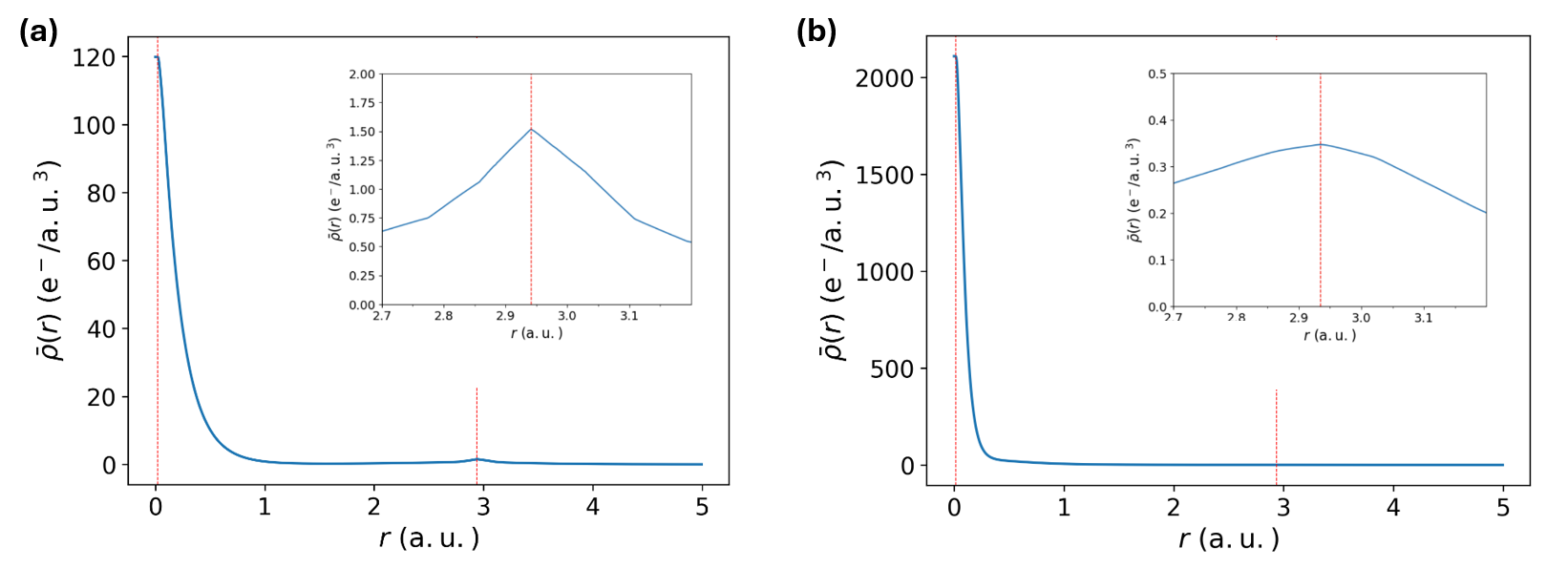}
\caption{\label{fig:LiF}
Sphericalized densities for the heteronuclear diatomic, LiF, at the equilibrium internuclear separation 1.564~\AA~(2.956 a.u.) (a) Origin of sphericalized density at Li. The location of the F atom is discernible as a peak in the distribution at the correct internuclear separation (see inset). (b) The second sphericalized density distribution, with F located at the origin. Although the peak for Li is smaller (due to its smaller size) it is still discernible in the inset, again at the correct internuclear separation.
}
\end{figure*}

\section{\label{sec:Num} Numerical examples: LiF and glycine}

We illustrate the encoding of atomic coordinate information within the spherical DFT electron density distributions using the LiF heteronuclear diatomic and the amino acid glycine (${\rm C}_2{\rm H}_5{\rm N}{\rm O}_2$). These systems serve to highlight three key features of spherical DFT in anticipation of practical applications to larger molecular systems: (i) the use of established distance geometry techniques to recover 3D spatial coordinates of constituent atomic nuclei; (ii) the persistence of formal nuclear (Kato) cusps\cite{Kato1957,Steiner1963,Handy1996,anderson2021non} even after spherically averaging the total Coulombic molecular density about individual nuclei; and (iii) the potential impact of computational noise in practical applications of the formal theory. Methods for mitigating (iii) have been extensively developed over the past decades, to enable the reconstruction of atomic coordinates from biomolecular and nanoscale electron densities in the presence of experimental noise, degenerate distance data deriving from an underlying symmetry, or a lack of complete information: these methods are discussed further below in Section~\ref{sec:Discussion}.

For both  LiF and glycine, the total electron density of the molecule at its optimized ground state geometry was first generated on a cubic grid with 0.0833 a.u.~spacing between points. The sphericalized electron density for each atom in the molecule was computed as a function of the radial distance $r$ from its center by angle-averaging about the center using a custom Python code implementing Eq.~(\ref{eq:spherical-den}). The sets of sphericalized electron densities were then used to reconstruct the original molecular geometry using distance geometry methods.

\subsubsection{Sphericalized density distributions for LiF}

The molecular electron density distribution for LiF was calculated at the MP2/cc-aug-pVQZ level of theory, at the experimental equilibrium internuclear separation of 1.564 \AA .\cite{NIST-LiF-2025}  For each atom of LiF, the sphericalized electron density was computed as a function of the radial distance \textit{r} from its nucleus, with the results shown in Fig.~\ref{fig:LiF}. Two sphericalized distributions are shown: one with Li located at the origin and the other with F at the origin. In both distributions, there is a discernible peak at a radius 1.564~\AA~ from the origin. 

Since LiF is a diatomic, the locations of the sphericalized peaks trivially determine the geometry of the molecule. Each atom's sphericalized density distribution locates the other atom 1.564~\AA~ away. Complexity arises when more than two atoms are involved in a three-dimensional space, as will be seen next for glycine.

\subsubsection{Sphericalized density distributions and atomic peak identification for glycine}
\begin{figure}
\includegraphics[width=0.8\linewidth]{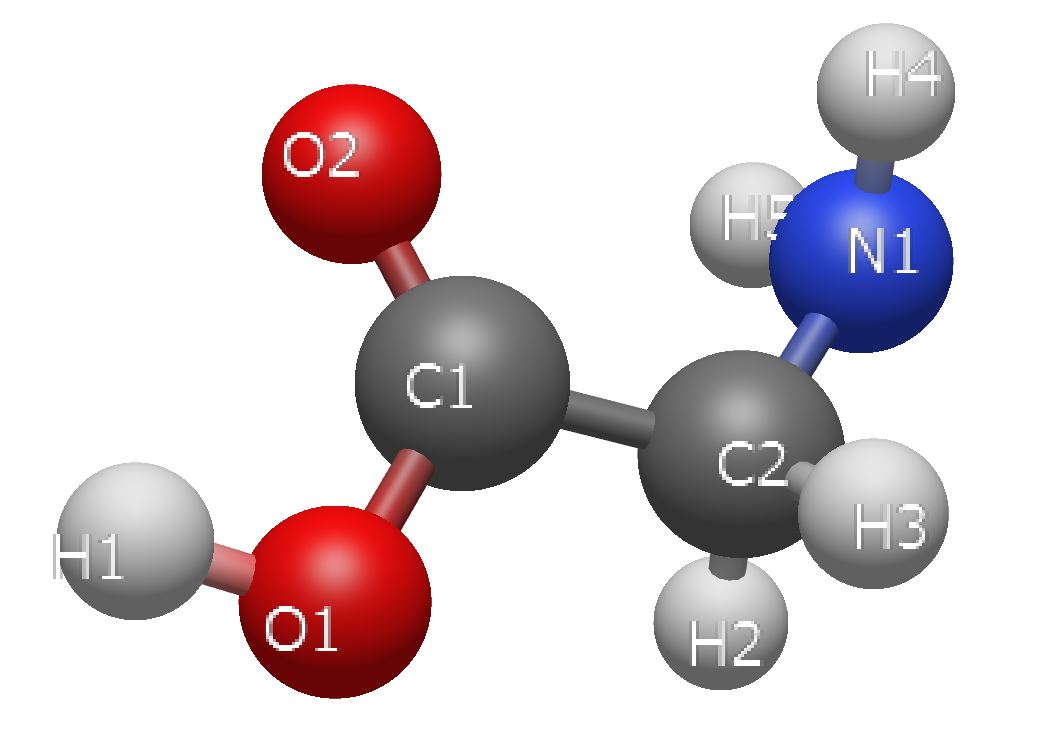}
\caption{\label{fig:glycine} Optimized geometry of glycine from quantum mechanical calculations. Converged atomic coordinates are given in Table~\ref{tab:opt-coords}.}
\end{figure}

\begin{table}
\caption{\label{tab:opt-coords} Glycine atomic coordinates (a.u.) following geometry optimization at MP2/cc-aug-pvQZ level of theory. Atom numbering corresponds to the designations used in the text; element labels for carbon, nitrogen, oxygen, and hydrogen correspond to those of Fig.~\ref{fig:glycine}.}
\begin{ruledtabular}
    \begin{tabular}{ccddd}
    Atom & Element& $x$ & $y$ & $z $ \\ \hline
    1 &    C1   & -1.020760  &  0.216286  & -0.000369 \\ 
    2 &    C2   &  1.363295  & -1.379247  & -0.000769 \\
    3 &    N1   &  3.715490  &  0.015918  &  0.000302 \\
    4 &    O1   & -3.111517  & -1.262518  &  0.000536 \\
    5 &    O2   & -1.097063  &  2.500065  & -0.000145 \\
    6 &    H1   & -4.544376  & -0.126168  &  0.000783 \\
    7 &    H2   & 1.280290   & -2.617584  & -1.654015 \\
    8 &    H3   & 1.279915   & -2.619449  &  1.651031 \\
    9 &    H4   & 3.795193   &  1.162414  &  1.533856 \\
   10 &    H5   & 3.793974   &  1.166756  & -1.530070 \\
    \end{tabular}
\end{ruledtabular}
\end{table}

\begin{figure}
\includegraphics[width=1\linewidth]{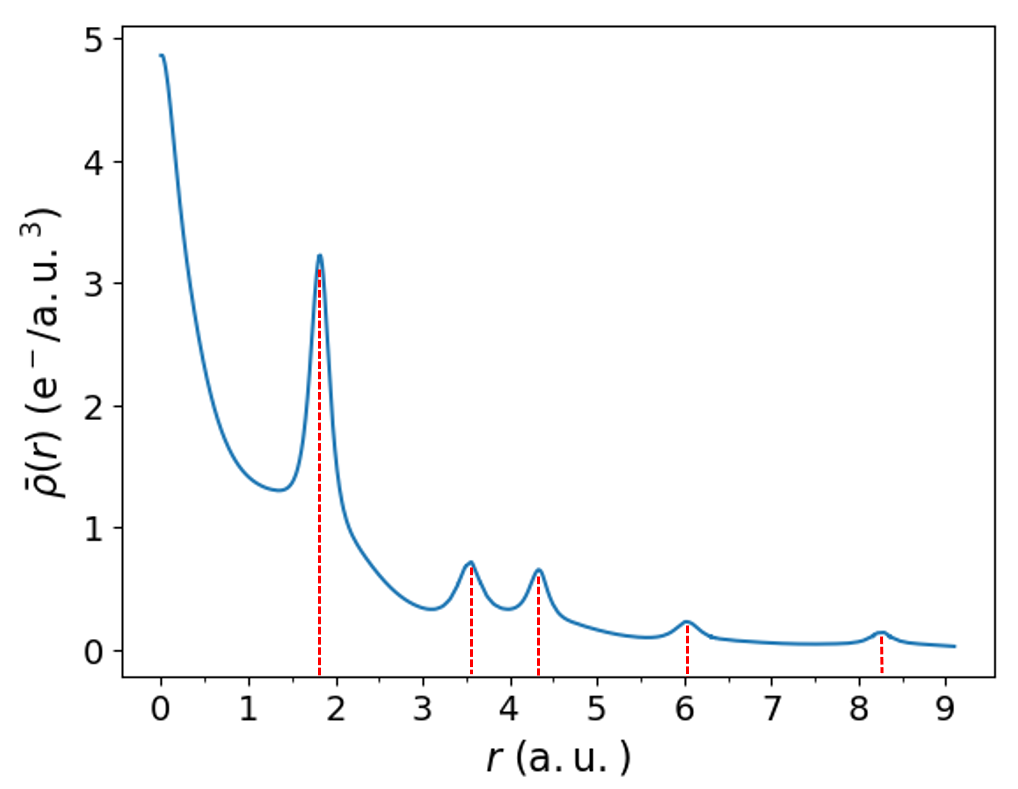}
\caption{\label{fig:atom6}Spherical density distribution as a function of radial distance from the nucleus of atom 6 of glycine (subsequently identified as a hydrogen atom). Although the electron density at the nucleus dominates the distribution, additional peaks are easily identified numerically as local maxima (indicated via red vertical lines), corresponding to the radial distances of other atoms from the center. As the distance from the origin increases, the magnitude of the peaks decreases, but they remain present.}
\end{figure}

Using starting coordinates for the lowest-energy conformer of glycine, designated Ip in the previous theoretical study by Csaszar, \cite{csaszar1992conformers} geometry optimization was performed using the Gaussian~16 electronic structure code\cite{g16} at the MP2/aug-cc-pVQZ level of theory. The optimized geometry is illustrated in Fig.~\ref{fig:glycine}, and the corresponding  atomic coordinates are listed in Table~\ref{tab:opt-coords}. These serve as the reference for comparison with the reconstructed atomic coordinates derived from the sphericalized atomic densities. 

\begin{figure*}
\includegraphics[width=1.0\linewidth]{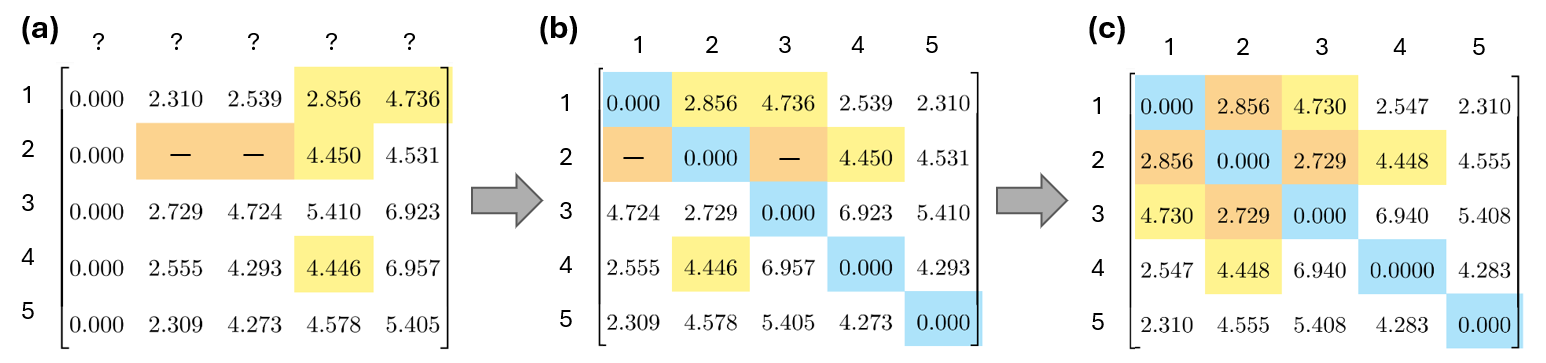}
\caption{\label{fig:matrixtrans}Matrix transformations to prepare the distance matrix for processing into coordinates by the MDS procedure. (a) Initial $5\times5$ matrix constructed from 10 sphericalized densities as described in the text. Peak measurements with larger uncertainties are highlighted in yellow. Orange-highlighted entries correspond to the overlapping peaks underneath the broad peak of atom 2 (see Fig.~\ref{fig:atom2}) as described in the text. Columns are labeled with `?' to indicate that they have not yet been reorganized to enforce approximate symmetry about the diagonal. (b) Reorganized raw distance value matrix enforcing approximate (within measurement error) symmetry about the diagonal. The columns are now labeled symmetrically to correspond to the atom ordering of the rows. The distance from any atom to itself is zero, comprising the diagonal of the distance matrix. (c) Finalized distance matrix for input to the MDS procedure. To enforce strong symmetry, values have been averaged across the diagonal. Missing values (orange) have been replaced by their symmetric counterparts.}
\end{figure*}

The total molecular electron density of the molecule at its optimized ground state geometry was used to generate the sphericalized densities about each of the ten glycine atoms using the custom Python code. The sphericalized density for atom 6 of glycine is illustrated in Fig.~\ref{fig:atom6}, with six nuclear cusp peaks identified, indicated by vertical red dashed lines. These correspond to the nuclear cusp at the central atom, a hydrogen, and five additional ``heavy atom'' (N,C,O) peaks of the glycine molecule. (Peak assignments to specific atoms emerge from the symmetrization procedure illustrated in Fig.~\ref{fig:matrixtrans}).
\begin{figure*}
\includegraphics[width=0.75\linewidth]{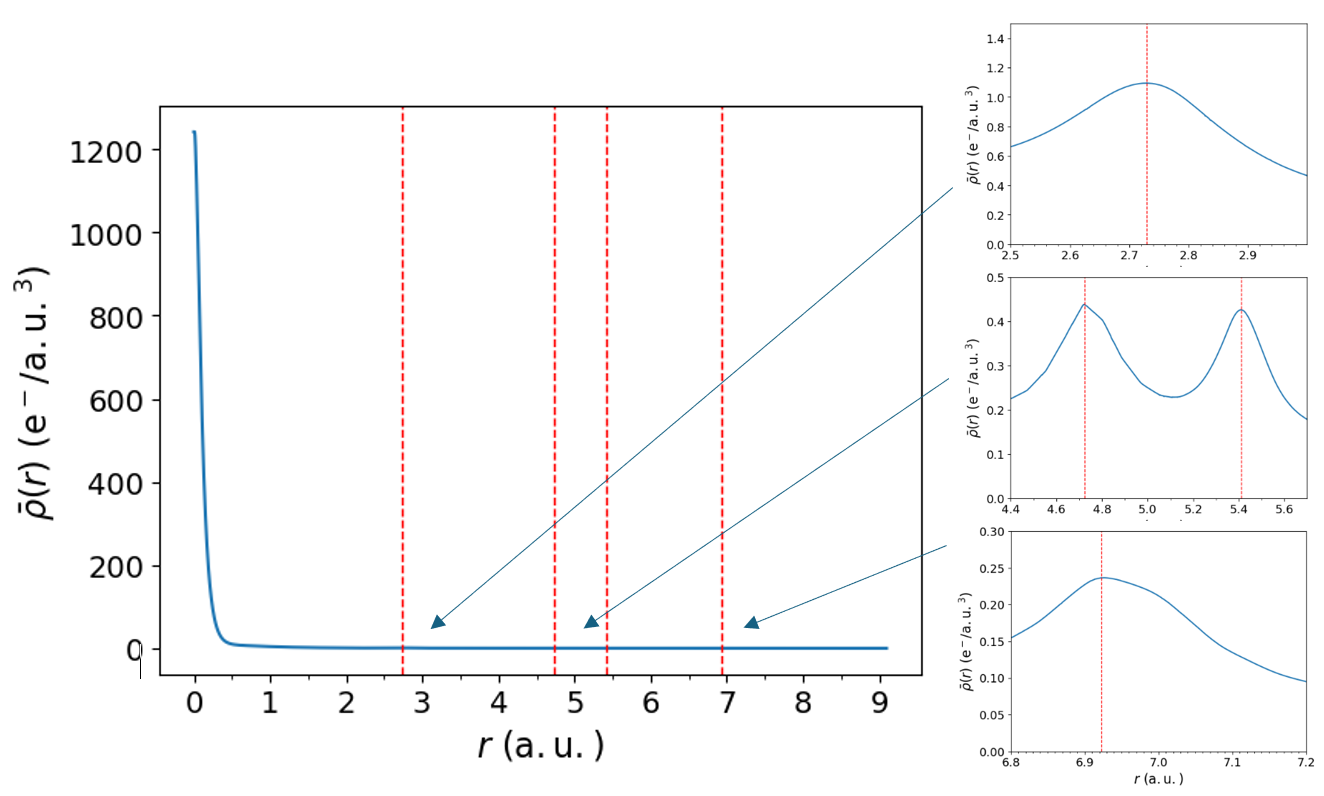}
\caption{\label{fig:atom3}Spherical density distribution as a function of radial distance from the nucleus of a heavier atom, atom 3 (subsequently identified as a nitrogen atom), of glycine. In a zoomed-out view, the density at the nucleus dominates the distribution. However, peaks are still detectable, and their form can be observed, when zooming in (see subplots). A total of five peaks are present, including the one at the origin. As described in the text, none of the hydrogen atoms of glycine are detectable.}
\end{figure*}

\begin{figure}
\includegraphics[width=0.9\linewidth]{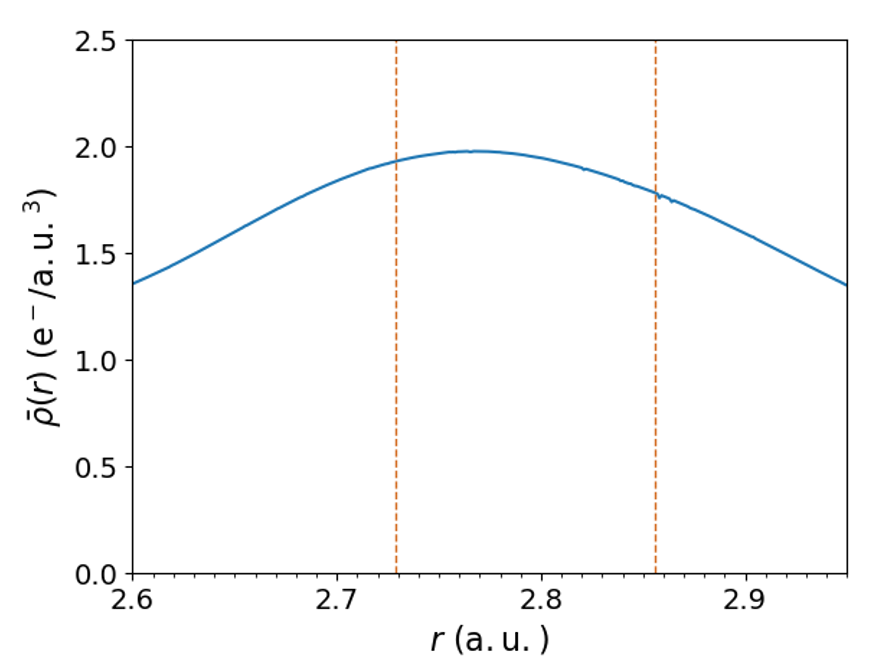}
\caption{Zoomed-in view of a combined peak in the sphericalized density distribution of atom 2 (subsequently identified as a carbon atom). Due to the geometry of glycine (see Fig.~\ref{fig:glycine}), atoms 1 and 3 are located at nearly identical radial distances away from atom 2. Thus, they are observed as a single, combined peak. Due to the inability to resolve their individual distances from atom 2, these two peak locations were labeled as missing values (see orange cells in Fig.~\ref{fig:matrixtrans}a). In the subsequent data analysis, they were replaced with their corresponding distance measurements from across the diagonal in the distance matrix (see Fig.~\ref{fig:matrixtrans}c). The replaced peak values are indicated by orange vertical lines in the plot.}
\label{fig:atom2} 
\end{figure}

In principle, each sphericalized density for a glycine atom should exhibit a total of ten peaks, corresponding to ten distances: nine from the detected peaks of other atoms, and a zero distance peak for the atom about which the sphericalization was performed. In the present case, however, only five or six peaks were detected per atom. The missing peaks were determined to correspond to the five hydrogen atoms of glycine, most likely washed out due to their lack of significant electron density and the large background densities contributed by other, heavier atoms in the molecule.  The spherical densities corresponding to the hydrogen atoms were precisely those for which six peaks could be detected, with the peak at the origin corresponding to the hydrogen itself, plus the five heavy atoms; see Fig.~\ref{fig:atom6}.  In one case (atom 2), only four peaks were detected; the broad observed peak (Fig.~\ref{fig:atom2}) corresponds to two overlapping peaks due to two atoms located at similar radial distances from the origin.

The set of atomic peak distances detected within each sphericalized density contributes a distinct atomic ``fingerprint'' of information from the given atom.  Analyzed together using distance geometry techniques, the combined set of spherical density fingerprints enables a global description of the molecular system, and specification of the atomic locations of all atoms in the molecule.  It is important to note that each calculated spherical density distribution exists as its own entity in isolation: once computed, it retains no knowledge of the atomic location about which its sphericalization was performed.  It is only when the distance information from all atoms in the molecule are analyzed in combination that the atomic center locations of the sphericalized densities can be derived.

\begin{figure*}
\includegraphics[width=0.65\linewidth]{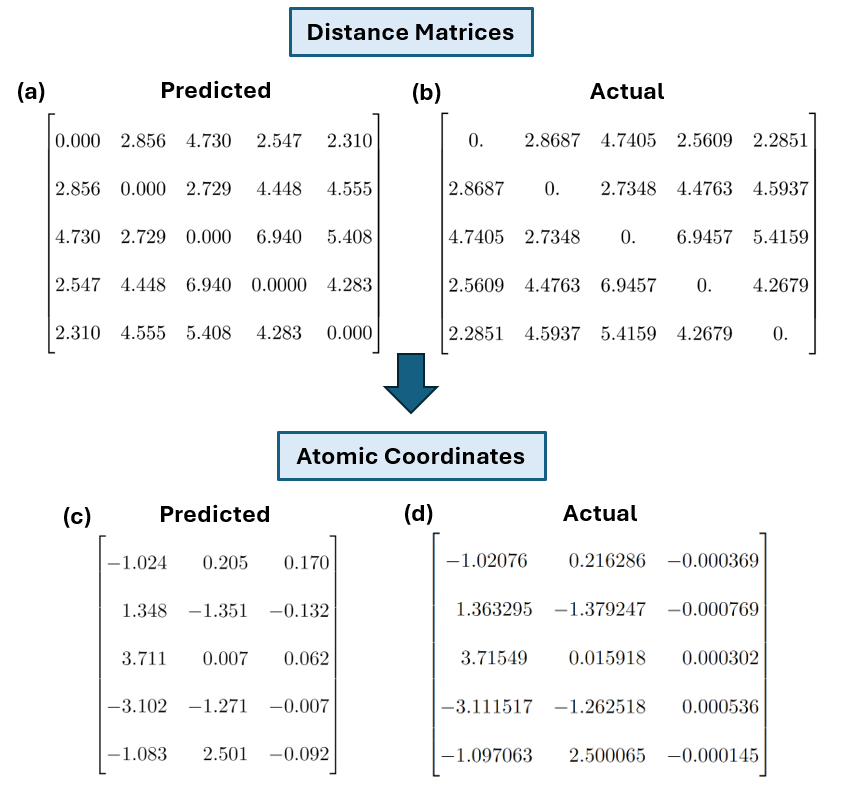}
\caption{Results of MDS analysis. The predicted $5\times5$ distance matrix (a) was constructed using peak identification and matrix transformations, as described in Fig. 4. The actual distance matrix for the corresponding 5 atoms, following structural optimization in Gaussian~16, is shown in (b). Using the MDS procedure with Procrustes alignment as described in Appendix~B, the glycine heavy-atom distance submatrix yields the predicted atomic $(x,y,z)$ coordinates (c). The actual atomic coordinates from a Gaussian~16 geometry optimization are given in (d).}
\label{fig:Matrixfin}
\end{figure*}

\begin{figure}
\includegraphics[width=0.82\linewidth]{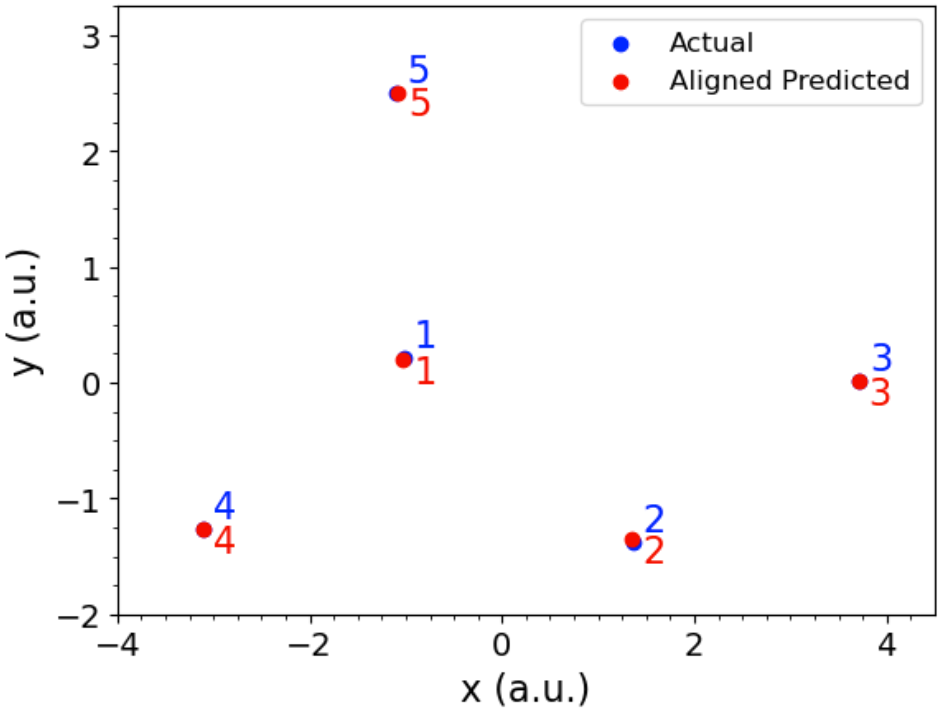}
\caption{Two-dimension MDS projection comparing actual coordinates (blue) to the procrustes-aligned predicted coordinates (red). Atoms are numbered according to the labels in Fig.~\ref{fig:matrixtrans}}
\label{fig:coordplot}
\end{figure}

\subsubsection{Inversion of distance information to extract 3D atomic coordinates}
For glycine, while the heavier atoms are detectable in the hydrogens' electron density distributions, the hydrogen atoms themselves were not discernible in the sphericalized electron density distributions of the heavier atoms.  Thus, only partial information on the atomic locations was recoverable. In order to take advantage of the multidimensional scaling (MDS) algorithm\cite{crippen1978stable,dokmanic2015euclidean} for the assigned distance geometry problem for purposes of illustration, we focus on analyzing the distance submatrix for the five heavy atoms (N, 2C, 2O) only. This matrix was constructed by extracting paired distance values ($d_{ij} = d_{ji}$ for atoms $i$ and $j$) from the initial $10\times 10$ matrix involving all atoms (including missing distance data for the hydrogen atoms). The resulting $5\times 5$ matrix is shown in Fig.~\ref{fig:matrixtrans}(a). The MDS algorithm is described in Appendix~B.

Application of the MDS algorithm to invert coordinates from distances requires a symmetric distance matrix with zero values along the diagonal.  The relevant matrix transformations are illustrated in Fig.~\ref{fig:matrixtrans}. First, approximate symmetry is imposed by reorganizing each row to align corresponding distances across the diagonal.  In the present instance, this could be accomplished manually due to the small atom count. The result is shown in Fig.~\ref{fig:matrixtrans}b. In practice, for larger systems, this operation is performed using an optimization routine.\cite{dokmanic2015euclidean}  Next, to enforce exact symmetry, each distance value is averaged with its partner value across the diagonal.  Missing values were replaced with those of their symmetric partners. This gave the final symmetric distance matrix of Fig.~\ref{fig:matrixtrans}c, to which the MDS procedure could now be applied.

The original optimized molecular coordinates calculated from Gaussian~16 were used as the reference for assessing the accuracy of the atomic coordinate reconstruction using the MDS algorithm. Since the MDS-generated coordinates are reported in an arbitrary coordinate system, a {\it Procrustes alignment procedure} \cite{dokmanic2015euclidean} (see Appendix~B) was used to rotate and translate the MDS-computed heavy-atom coordinates for comparison with the quantum mechanical reference values. MDS and Procrustes procedures from \texttt{SciPy}\cite{2020SciPy-NMeth} were used. The results are shown in Figs.~\ref{fig:Matrixfin}c and \ref{fig:Matrixfin}d.

As is apparent from the coordinates listed in Fig.~\ref{fig:Matrixfin}d, the heavier (non-hydrogen) atoms of glycine lie close to the $x-y$ plane (indicated by small magnitudes of their $z$ coordinates). This offers an explanation for why the predicted coordinates are least accurate in placing the atoms correctly along the z-axis (see Fig.~\ref{fig:Matrixfin}c), despite the similarity between the distance matrices for the predicted and actual data (see Figs.~\ref{fig:Matrixfin}a and \ref{fig:Matrixfin}b).  As a second numerical experiment, the MDS algorithm was rerun to project the final coordinates onto a 2D plane (corresponding to the $k=2$ setting of the MDS algorithm.) The predicted and actual coordinates are compared on a 2D plot in Fig.~\ref{fig:coordplot}. In this representation, the close agreement between predicted and actual coordinates is evident.

\section{Discussion and conclusions\label{sec:Discussion}}
We have presented an extension to the proofs of Theophilou\cite{theophilou2018} and Nagy\cite{nagy2018} establishing spherical DFT for Coulombic systems, demonstrating that {\it a priori} knowledge of the atomic center locations in a molecule is not required in order to uniquely determine the external potential from the set of constituent sphericalized densities about each atom.  This result follows from the theorems of distance geometry, given knowledge of all $M(M-1)/2$ unique pairwise distances between atoms, as determined from the nuclear cusps appearing in the sphericalized density set.  This allows the construction of a corresponding distance matrix,\cite{dokmanic2015euclidean} and the unique determination of the sphericalized density coordinate locations as demonstrated here using multidimensional scaling analysis.  It follows that $\{\bar{\rho}_i(r_i)\}, i=1,\ldots M \ \Rightarrow\ v({\bf r})$: the equivalent of the Hohenberg-Kohn theorem for Coulombic systems in spherical DFT.  

As noted by Nagy, spherical DFT can also be formulated as a constrained-search problem, generalizing the theory to non-$v$-representable densities and non-Coulombic potentials.\cite{nagy2018}  In principle, this may allow future extensions of spherical DFT to more general problems such as a molecule in an applied field or a molecule in an effective field induced by coupling to a solvent; non-Coulombic potentials that are sufficiently singular as to enable distance determinations to other centers; or potentials with multiple singularities at the same center. For application to molecular systems, the constrained search procedure requires the identification of a parent many-electron wavefunction $\Psi$ that yields the set of sphericalized densities at all $M$ nuclei.  If the nuclear locations are unknown, one possibility is to construct an iterative energy minimization scheme starting from an initial guessed set of nuclear locations that would allow the discovery of a unique parent wavefunction and total electron density associated with a given spherical density set. (The identification of a density whose sphericalizations give rise to a given set of $\{\bar{\rho}_i(r_i)\}$ is referred to as the "set-representability problem."\cite{nagy2021spherical}) However, without knowledge of the originating nuclear locations to serve as registration points, the search space for the energy minimization will grow exponentially with the number of atoms in the system, due to the self-consistency requirement for the collection of sphericalized densities and the total density from which those densities were derived.  In this respect, for molecular systems, the present work can be seen as effecting a projection from that exponentially large search space onto a much smaller subspace constrained by information already implicit in the assembly of sphericalized densities.

The numerical examples for LiF and glycine provide concrete illustrations of how the nuclear cusp peaks, and thus, internuclear distance information, remain intact within the set of sphericalized densities for a given molecule.  As expected, the magnitudes of the cusp peaks decrease as a function of distance from a given origin atom (see, {\it e.g.},~Fig.~\ref{fig:atom6}), since densities at greater radial distances are subject to angle-averaging over a larger spherical surface area.  The use of differentiable Gaussian basis functions to describe the electron density results in numerical smoothing of the nuclear cusp peaks; nevertheless, their signatures survive.  The combination of this effect with the overall intensity decay as a function of distance from the origin atom is the likely explanation for our inability to identify hydrogen atom peaks in densities sphericalized about the C, N, and O atoms in glycine. While the primary focus in these examples is to illustrate the formal theory, the issues of missing or uncertain distance data can be addressed in practical applications via techniques developed for the unassigned distance geometry problem, where the input data is provided as an unsorted list of distances.\cite{dokmanic2015euclidean,duxbury2016unassigned,billinge2018recent} 

A related question concerns how nuclear coordinate locations can be identified in high-symmetry molecules with  degenerate distance data. In such systems, multiple atoms in a molecule will 
share a common radius about a given origin, resulting in a combined peak in the spherically-averaged profile.  
From a purely theoretical standpoint, the number of pairwise distances is still $N(N-1)/2$ for $N$ atoms, notwithstanding the degeneracy in atomic locations (and thus degenerate values in the distance matrix) induced by the molecular symmetry.
Consequently, the distance geometry theorem\cite{havel1983theory} discussed in Section~\ref{sec:Encoding}.B still holds.  In practice, however, degenerate distance data can arise from an exact atomic symmetry of the system as in C$_{60}$, or an accidental (near)-degeneracy as observed in Fig.~\ref{fig:atom2} for glycine. In addition, experimental distance data may be incomplete or noisy.  Methods for reconstructing 3D protein coordinates from incomplete or degenerate NMR distance data associated with atomic pairs (the \textit{assigned} distance geometry problem) typically rely upon the imposition of additional geometric and energetic constraints.\cite{crippen1977novel,crippen1978stable,havel1984distance,billinge2016assigned,liberti2014euclidean}  For the {\it unassigned} distance geometry problem mentioned above, researchers have successfully utilized stochastic heuristics such as genetic algorithms, the Liga build-up algorithm with backtracking, and other methods based upon random structure generation with refinement, to reconstruct nanostructures from pair distribution function (PDF) data.\cite{juhas2006ab,duxbury2016unassigned,billinge2016assigned,billinge2018recent,terban2022structural} 

In addition to providing new insight into the fundamental theorems of density functional theory, spherical DFT has inspired discussion of potential connections to the atom-in-molecule problem.\cite{theophilou2018,nagy2018,nagy2019}  The atom-in-molecule densities $\rho_i^*({\bf r})$ in an $M$-atom molecule satisfy the relation
\begin{equation}
    \rho({\bf r}) = \sum_{i=1}^M \rho_i^*({\bf r})
\end{equation}
at every point ${\bf r}$ in space. Clearly, the $\{\rho_i^*({\bf r})\}$ are not unique, and for a bound molecular system, they are not perfectly spherical.  Each $\rho_i^*({\bf r})$ integrates to a number of electrons $N_i^*$ (not necessarily an integer), yielding an effective atomic charge $q_i = Z_i - N_i^*$ on the $i$th atom. The $q_i$ satisfy the constraint 
\begin{equation}
    q = \sum_{i=1}^M q_i,
\end{equation}
where $q$ is the overall charge on the molecule; $q = 0$ for a neutral system.

\begin{figure}
    \centering
    \includegraphics[width=1.0\linewidth]{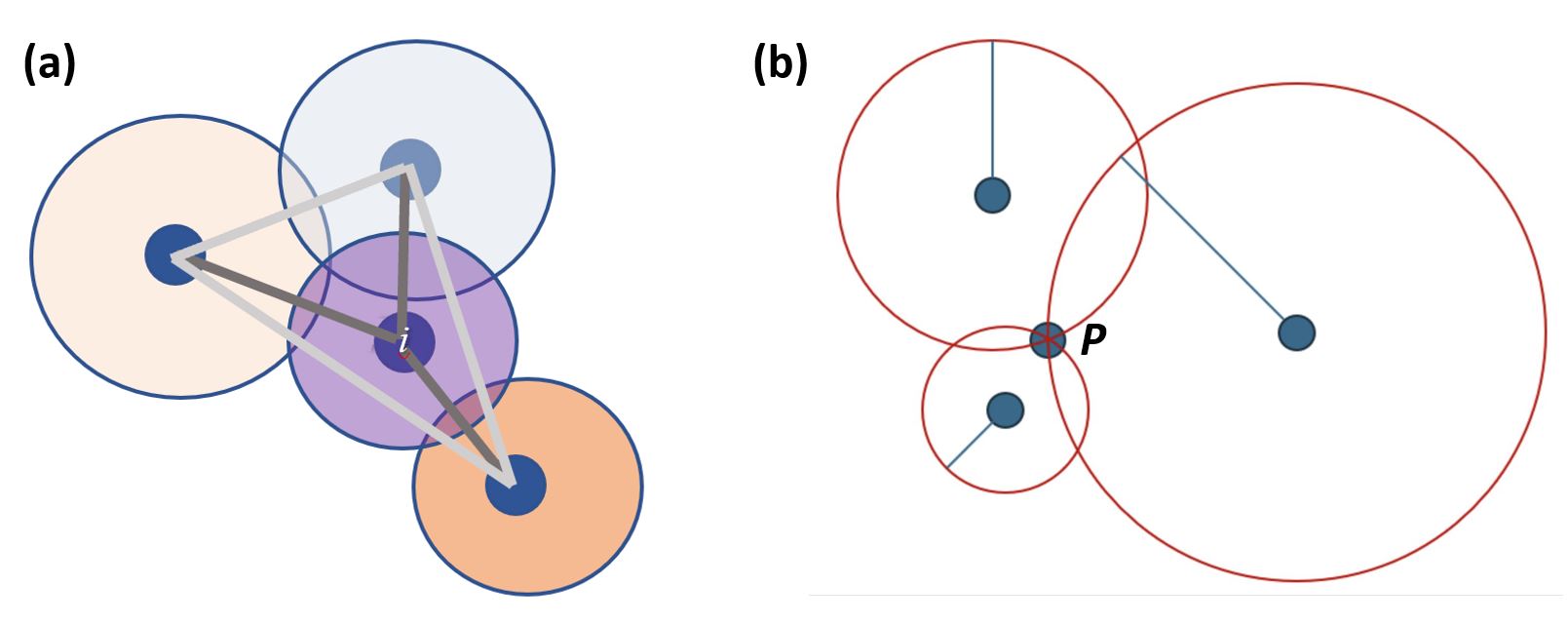}
    \caption{Interacting atoms with sphericalized densities. (a) Distance geometry perspective. Distances measured from  the central atom $i$ to neighboring nuclei (dark grey lines), coupled with distances measured between surrounding nuclei (light grey lines) provide a complete representation of relative spatial positioning. 
    (b) Schematic 2D trilateration ("GPS") perspective. The intersection of the three circles suffices to determine the location of atom $P$.}
    \label{fig:fuzzy}
\end{figure}

Since each spherical DFT density in a molecule integrates to the {\it total} number of electrons $N$ in the molecule, and together do not sum to $\rho({\bf r})$, the sphericalized densities do not correspond to proper atom-in-molecule densities.  Nevertheless, these "fuzzy atoms"\cite{fn2} (nuclei + sphericalized densities) embody essential information on the electronic structure and local atomic configurations characterizing an atom's chemical environment---analogous to the information inherent in an atom-in-molecule decomposition\cite{bader1990,parr2005} but without explicit angular orientation dependencies (see Fig.~\ref{fig:fuzzy}). By construction, the sphericalized densities---like true atomic densities---are spherically symmetric at both short- and long-range, and exhibit formal exponential behavior in both limits (see Eqs.~(\ref{eq:Kato2}) and (\ref{eq:long-range})).  Viewed from afar, the electron density of the molecule as a whole will resemble that of a localized, fuzzy, sphericalized atom, likewise integrating to $N$ electrons and decaying exponentially as $r \rightarrow \infty$.  An early connection between a fuzzy, atom-centered representation and the total electron density distribution was made by Becke, in his method for integrating electron densities in polyatomic molecules;\cite{becke1988multicenter} this algorithm is widely used in contemporary electronic structure codes. In spherical DFT, for Couombic systems, configurational information and complex bonding patterns survive angular averaging, and the sphericalized densities contribute collectively through their relative spatial center locations to yield a rigorous description of the total electron density of the system. 

A further connection to the atom-in-molecule can be seen in the matrix 
of pairwise cusp-to-cusp distances derived from the sphericalized electron density distributions 
({\it cf.}~Figs.~\ref{fig:matrixtrans} and \ref{fig:Matrixfin}), which recall the localization-delocalization matrices (LDMs) \cite{fradera1999lewis,matta2014modeling,koch2024analysis} of Bader's quantum theory of the atom-in-molecule (QTAIM).\cite{bader1985,bader1990} The LDMs are square matrices consisting of intra-atomic localization\cite{bader1990} and interatomic delocalization\cite{fradera1999lewis} indices computed from the molecule's QTAIM atomic basins, and they define a ``fuzzy'' molecular graph of the 
system.\cite{matta2014modeling,koch2024analysis}  
Although quite different in the types of invariants that define their respective matrix elements, the distance matrices of the present work and the QTAIM LDMs share a conceptual connection:
in both cases, pairwise information computed between atomic-like entities in a molecule is encoding information associated with the larger molecular system and associated molecular graph. Importantly, the accurate encoding of the molecular information in both formulations requires the use of true quantum mechanical densities rather than approximate forms.

In light of the extension demonstrated here, spherical DFT can be regarded as a formal \textit{descriptor theory} or \textit{generalized density functional theory},\cite{higuchi2004arbitrary,ayers2007alternatives} allowing the recasting of the electronic structure problem for a molecular system in terms of the set of sphericalized electron densities, and replacing the total density as the fundamental variable(s) of the theory. This situates spherical DFT in the company of other formally-established molecular descriptors, such as the density per particle (shape density);\cite{ayers2000density} the local kinetic energy and local electronic temperature;\cite{ayers2007alternatives} linear combinations of Kohn-Sham orbital densities;\cite{nagy2008alternative} the Shannon entropy density;\cite{nagy2013shannon} and Fisher information.\cite{nagy2025fisher} Establishing novel descriptors is important since they can potentially describe the properties of an interacting electronic system more efficiently than the total electron density; suggest new approaches for characterizing chemical bonding and reactivity; and facilitate the development of novel exchange-correlation density functionals.

Finally, spherical DFT provides a conceptual foundation for utilizing spherical atom-in-molecule densities to construct classical potential models of molecular structure based on interatomic distance information alone,\cite{Atlas2021} rather than modeling angular orientations directly, as is done in constructing machine-learned electron density distributions \cite{ghasemi2015interatomic,grisafi2019transferable,willatt2019atom,lee2022predicting} and potentials.\cite{thompson2015spectral,smith2017,behler2021four,takamoto2022b,kovacs2023evaluation}  The question of how to scale the description of atomic interactions in potential models to large, complex systems without requiring a combinatorically-large training dataset of angle- and orientation-dependent bonding exemplars represents an ongoing challenge in the design of transferable classical potentials applicable across diverse chemical systems. The ability to characterize densities and potentials using purely distance-dependent, spherical density-based representations of atomic environments suggests the possibility of using these descriptors as atom-centered latent variables in the design of chemically-aware scalable potentials.\cite{Atlas2021}  A recent example of how this approach can work in practice is the third generation of the AlphaFold neural network model for biomolecular structure prediction,\cite{jumper2021highly} for which the previous AF2 model was redesigned and extended to apply to more general chemical environments, while simultaneously eliminating the need for input angular information and equivariance considerations.\cite{abramson2024accurate}  In this sense, spherical DFT and its distance geometry foundation provide an elegant link between the NMR protein structure determinations pioneered over forty years ago, and the machine learning methods for molecular structure prediction of today.

\section*{Acknowledgements} We thank the UNM Center for Advanced Research Computing, supported in part by the National Science Foundation and NSF grant \# OCI-1040530, for providing the computational resources used in this work. S.~Samuels gratefully acknowledges internship support from NSF REU grant \# PHY-1659618 and support from the UNM Rayburn Reaching Up Fund.  The authors thank the anonymous reviewers for their valuable comments and suggestions.

\appendix
\renewcommand{\theequation}{A\arabic{equation}}\setcounter{equation}{0}
\section*{\label{sec:Appendix A} Appendix A. Evaluation of ${\cal I}_B$ in Eq.~(\ref{eq:AppA-result})}
We show that
\begin{equation}
    {\cal I}_B = 4 \pi \rho_0^B(r')f_{\rm B}(r') + \sum_{l>0} \rho_l^B(r')\, f_{\rm B}(r') \times {\rm poly}(r,r',R)|_{r' \ne 0},
    \label{eq:calIB}
\end{equation}
where ${\rm poly}(r,r',R)|_{r'\ne 0}$ denotes a polynomial in the variables $r$, $r',$ and $R$, with $r' \ne 0$.

Referring to the geometry defined in Fig.~\ref{fig:geom}, we see that
\begin{align}
    x' &= x \nonumber \\
    y' &= y \nonumber \\
    z' &= r \cos \theta - R  &{\rm if}\ r\cos \theta > R \nonumber \\
       &= r \cos \theta + R = -(r\cos \theta - R)  &{\rm if}\ r\cos \theta < R
\end{align}
We consider the case where $r\cos \theta > R$; the analysis for $r\cos \theta < R$ proceeds similarly.

Since $x^2 + y^2 = r^2 - z^2$ and $z^2 = r^2 \cos^2 \theta$, 
\begin{align}
    r'^2 &= x'^2 + y'^2 + z'^2 = x^2 + y^2 + (r\cos \theta - R)^2 \nonumber \\
    &= r^2 - 2rR\cos \theta + R^2.
\end{align}
Rewriting ${\cal I}_{\rm B}$ in terms of a spherical harmonic expansion about center B in coordinate system ($r'$,$\theta'$,$\phi'$):\cite{theophilou2018}
\begin{align}
{\cal I}_{\rm B} &= \sum_{l,m} \rho_{lm}^B(r')\, f_{\rm B}(r') \int Y_{lm}(\theta',\phi') \sin \theta d\theta d\phi \nonumber \\
                 &\equiv {\cal I}_B^{(1)} + {\cal I}_B^{(2)},
\end{align}
where we have separated out the $l=0$ term from the sum in the first line, defining
\begin{equation}
    {\cal I}_B^{(1)} \equiv \int \rho_0^B(r')\, f_{\rm B}(r') \sin \theta d\theta d\phi 
    \end{equation}
and
\begin{align}
{\cal I}_B^{(2)} &\equiv \sum_{l>0;m} \rho_{lm}^B(r') f_{\rm B}(r')\, \int Y_{lm}(\theta',\phi') \sin \theta d\theta d\phi \nonumber \\
&= \sum_{l>0;m} \rho_{lm}^B(r') f_{\rm B}(r')\, J(\theta' \phi'),
\label{eq:IB2}
\end{align}
with
\begin{equation}
    J(\theta',\phi') \equiv \int Y_{lm}(\theta',\phi') \sin \theta d\theta d\phi.
    \label{eq:Jdef}
\end{equation}
Note that the special case of $r = R$, corresponding to $r' = 0$ and the location of the nuclear cusp at atom B, is covered by the first term in Eq.~(\ref{eq:calIB}).
${\cal I}_B^{(1)}$ is trivial to compute, since the spherically-symmetric function $\rho_B^0(r')$ is independent of the integration angles $\{\theta, \phi\}$ associated with center A.  Thus,
\begin{equation}
    {\cal I}_B^{(1)} = 4\pi \rho_0^B(r')\, f_{\rm B}(r').
\end{equation}

Now consider the integral in Eq.~(\ref{eq:Jdef}).  Since $x' = x$ and $y' = y$ due to the choice of coordinate system for this two-center problem, we have that $\phi' = \phi$. Thus, $Y_{lm}(\theta',\phi') = Y_{lm}(\theta',\phi) \propto P_{lm}(\cos \theta')e^{im\phi}$, where $P_{lm}$ is the Legendre polynomial of degree $l$ and order $m$.\cite{messiah1976quantum} 
The integrals over $\theta'$ and $\phi$ are separable, and $\int_0^{2\pi} e^{im\phi} d\phi = 2\pi \delta_{m0}$, so we may write:
\begin{equation}
    J_l(\theta',\phi') = 2\pi \left ( \frac{2l+1}{4\pi}\right )^{1/2} \int_0^\pi P_l(\cos \theta') \sin \theta d\theta.
\end{equation}
Since $r\cos \theta > R$, we have
\begin{equation}
    z' = z - R = r\cos \theta - R = r'\cos \theta'.
\end{equation}
We now change variables, setting $u=\cos \theta'$. Then
\begin{equation}
    u = \cos \theta' = \frac{r\cos \theta - R}{r'},
\end{equation}
\begin{equation}
    du = -\frac{r}{r'} \sin \theta d \theta ,
\end{equation}
and
\begin{equation}
\sin \theta d\theta = -\frac{r'}{r} du.
\end{equation}
This gives:
\begin{equation}
    J_l =  \int_{\frac{r-R}{r'}}^{-\frac{r+R}{r'}} \left (-\frac{r}{r'}\right ) P_l(u) du.
    \label{eq:Jeval}
\end{equation}
Note that since $r\cos \theta > R$, $r'$ is always $> 0$ and the integral in Eq.~(\ref{eq:Jeval}) is defined.  Since $P_l(u)$ is a polynomial in $r$, $R$, and $r'$, so is $J_l$.  Substituting into Eq.~(\ref{eq:IB2}), we obtain the result in Eq.~(\ref{eq:calIB}) as claimed.

\appendix
\renewcommand{\theequation}{B\arabic{equation}}\setcounter{equation}{0}
\section*{\label{sec:Appendix B} Appendix B. The Multidimensional Scaling (MDS) Algorithm} 

Starting from a distance matrix, classical multidimensional scaling (MDS) can be used to reconstruct coordinate locations, projected onto a $k$-dimensional space.  This process is outlined in Refs.~\onlinecite{crippen1978stable} and \onlinecite{dokmanic2015euclidean}. Here we follow the notation and presentation of [\onlinecite{dokmanic2015euclidean}].

We are given an $n\times n$ distance matrix $D$ with entries $d_{ij}$ corresponding to the distance between atoms $i$ and $j$. $n$ is the number of atoms in the system.  To prepare the distance matrix for eigenvalue decomposition, it is first transformed using an $n \times n$ \textit{geometric centering matrix}, $J$:
\begin{equation}
J = I - \frac{1}{n} \mathbf{1}\mathbf{1}^T, 
\end{equation}
where $\mathbf{1}\mathbf{1}^T$ denotes the $n \times n$ ones matrix ($\mathbf{1}$ is the $n \times 1$ vector of all ones).  Using $J$, the {\it Gram matrix} $G$ is computed as: 
\begin{equation}
G = -\frac{1}{2} J D^2 J
\end{equation}
$G$ represents the inner product relationships between the points in the lower ($k$)-dimensional space. Next, perform an eigenvalue decomposition of \( G \):
\begin{equation}
G = V \Lambda V^T,
\end{equation}
where \( V \) are the eigenvectors and \( \Lambda \) is the diagonal matrix of eigenvalues.  Select the top \( k \) eigenvalues in absolute value and their corresponding eigenvectors; \( k \) corresponds to the dimensionality of the coordinates in the lower-dimensional space. For example, the top three eigenvalues and corresponding eigenvectors would be chosen for a target 3D coordinate representation. 

The distance matrix-derived coordinates are then given by the $n \times k$ matrix $X$:
\begin{equation}
X = V_k \Lambda_k^{1/2}
\end{equation}
where \( V_k \) and \( \Lambda_k \) are composed from the eigenvectors and eigenvalues of the top \( k \) components.  These coordinates correspond to the $k$-dimensional representation of the original points.

Note that the MDS algorithm outputs coordinates in an arbitrary basis.  If comparison with a known set of coordinates is desired, an alignment procedure must be performed, as described in detail in [\onlinecite{dokmanic2015euclidean}] and summarized here.
Given two sets of coordinates $P$ and $Q$ of equal length $n$, but specified in two different coordinate systems, we first center both sets about their respective origins by subtracting their mean position values \(\mathbf{\bar{p}} \) and \(\mathbf{\bar{q}}\), from all other coordinates in each set, to form the centered sets $\bar{P}$ and $\bar{Q}$:
\begin{equation}
    \bar{P}_i = P_i - \mathbf{\bar{p}}  \quad \bar{Q}_i = Q_i - \mathbf{\bar{q}},\ \ 1 \le i \le n,
\end{equation}
where
\begin{align}
    \mathbf{\bar{p}} = (\bar{x}_p, \bar{y}_p, \bar{z}_p) \quad
    \mathbf{\bar{q}} = (\bar{x}_q, \bar{y}_q, \bar{z}_q)
\end{align}
Next, an \textit{orthogonal Procrustes analysis} is applied to the centered sets in order to minimize the difference between $\bar{P}$ and $\bar{Q}$. This involves finding the matrix $R$ that best maps $\bar{P}$ onto $\bar{Q}$ via (rigid) rotation and reflection transformations: 
\begin{equation}
R = \argmin_{A:\ A^T A = I} \|A\bar{P} - \bar{Q}\|_F^2,
\label{eq:Rdef}
\end{equation}
where $|| \cdot ||_F$ is the Frobenius norm.  The solution $R$ to Eq.~(\ref{eq:Rdef}) is expressed in terms of the singular value decomposition (SVD) of $\bar{P}^T \bar{Q}$:
\begin{align}
    \bar{P}^T \bar{Q} &= U \Sigma V^T \nonumber \\
    R &= V U^T.
\end{align}
The optimal rotation matrix $R$ can be used to map the point set $P$ onto $Q$, but since the mean points of both sets were subtracted for centering purposes, they must be added back in order to maintain correct positioning. The final transformation applied to $\bar{P}$ to calculate the aligned point set $P_{\text{aligned}}$ is:
\begin{equation}
P_{\text{aligned}} = R(\bar{P} - \mathbf{\bar{p}}\mathbf{1}^T) + \mathbf{\bar{q}}\mathbf{1}^T.
\end{equation}

\section*{\label{sec:Appendix C} Appendix C. Asymptotic Decay of the Sphericalized Densities} 
At long range, true atomic and molecular densities will decay exponentially with an exponent related to the least-negative occupied orbital energy $\epsilon_{I}$ (Eq.~(\ref{eq:long-range})). According to the extended Koopmans' theorem,\cite{smith1975extension,morrell1975,ernzerhof2009} $\epsilon_{I}$ for an atom or molecule provides a good estimate of the first ionization potential.

\begin{figure*}
\includegraphics[width=1.0\linewidth]{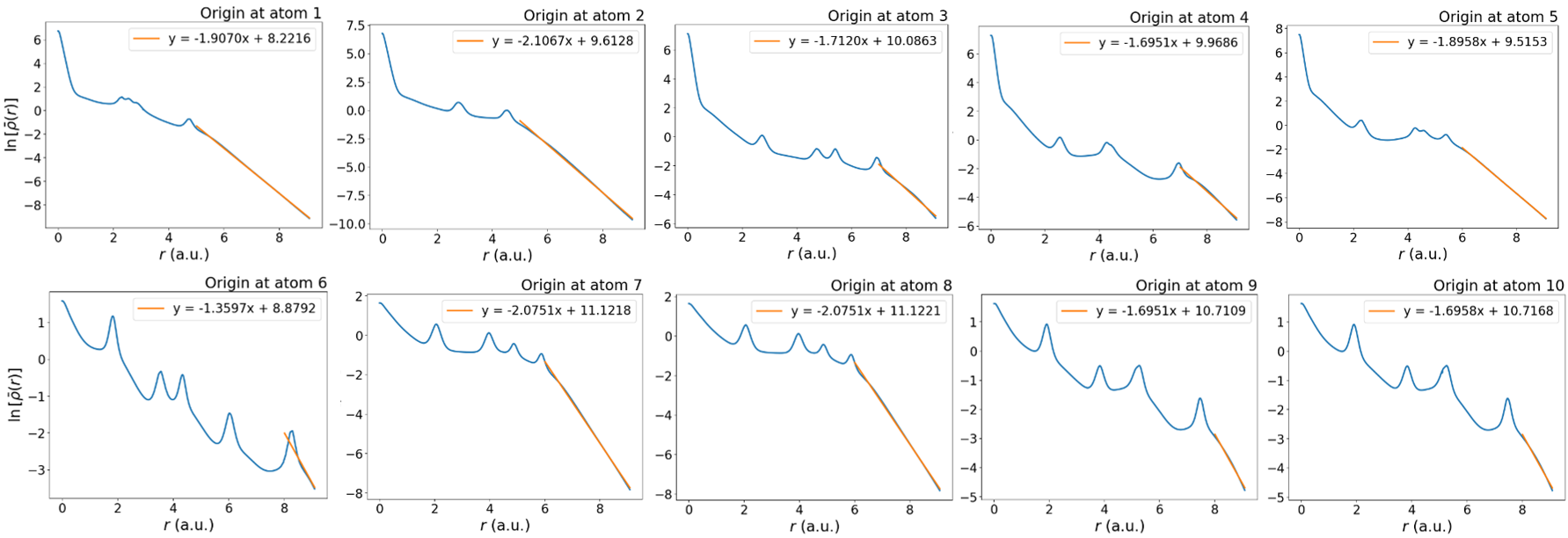}
\caption{Sphericalized electron densities for each atom of glycine, plotted on a natural log scale with units of $\ln(\mathrm{e^-}/\mathrm{a.\!u.\!^3})$. The long-range exponential decay is fitted to a line to determine the slope (orange). The $\ln$ scale enhances identification of the Kato cusps.}
\label{fig:LinearFits} 
\end{figure*}

To explore how well this relationship holds for the sphericalized densities, we plotted the natural log of each sphericalized density as a function of radial distance. The results for all 10 atoms of glycine are shown in Fig.~\ref{fig:LinearFits}. The log plots incidentially emphasize the cusp locations of other atoms: compare Fig.~\ref{fig:atom3} to Fig.~\ref{fig:LinearFits} for atom 3. 

We determined the slope via a linear regression at long range in each of the natural log plots. The slopes were then converted to predicted ionization energies. The results are given in Table \ref{tab:ionization_energy}.

\begin{figure}
\includegraphics[width=1.0\linewidth]{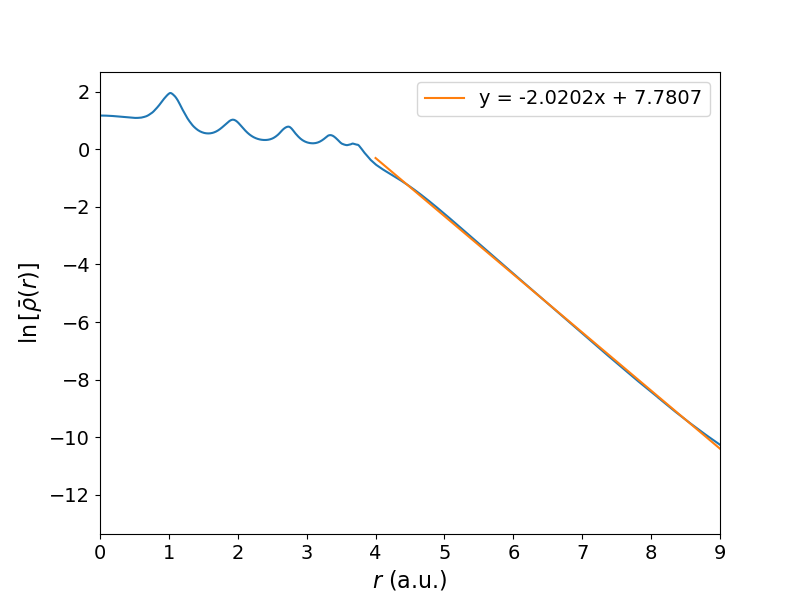}
\caption{Sphericalized electron density computed with respect to glycine's nuclear charge center of mass, plotted on a natural log scale. The long-range exponential decay is fitted to a line to determine the slope.}
\label{fig:GlycineCOM}
\end{figure}

\begin{table}[htbp!]
\centering
\caption{Results from exponential fits to sphericalized density data for glycine, and resulting estimated ionization potential values.\ c.m. denotes the atomic number center of mass of the molecule.}
\begin{ruledtabular}
\begin{tabular}{c|c|c|c|c}
Atom & Radial fitting & Slope from & Ionization & Ionization \\ 
Number & range (a.u.) & fit (1/a.u.) & energy (a.u.) & energy (eV) \\ \hline
1  & 5.0--9.0 & $-1.9070$ & 0.4546 & 12.3703 \\ 
2  & 5.0--9.0 & $-2.1067$ & 0.5548 & 15.0955 \\ 
3  & 7.0--9.0 & $-1.712$0 & 0.3664 & 9.9697  \\ 
4  & 7.0--9.0 & $-1.6951$ & 0.3592 & 9.7734  \\ 
5  & 6.0--9.0 & $-1.8958$ & 0.4493 & 12.2251 \\ 
6  & 8.0--9.0 & $-1.3597$ & 0.2311 & 6.2889  \\ 
7  & 6.0--9.0 & $-2.0751$ & 0.5382 & 14.6465 \\ 
8  & 6.0--9.0 & $-2.0751$ & 0.5382 & 14.6462 \\ 
9  & 8.0--9.0 & $-1.6951$ & 0.3592 & 9.7730  \\ 
10 & 8.0--9.0 & $-1.6958$ & 0.3595 & 9.7818  \\
c.m. & 4.0--9.0 & $-2.0202$ & 0.5101 & 13.8813 \\
\end{tabular}
\end{ruledtabular}
\label{tab:ionization_energy}
\end{table}

We find predicted ionization energies in a similar range as the experimental and theoretical values compiled by NIST, 8.8--10 eV,\cite{NIST-glycine} and the recent consensus value of 10.0 eV given by de Souza and Peterson,\cite{de2022high} based on a survey of the experimental and theoretical literature. The variation observed here across atoms is due to the measurements being performed relatively close to the respective atomic centers, resulting in inconsistent quality across the linear fits.  As an example, for atom 6, the range over which the linear fit could be performed was limited by the size of cube on which the molecular density was generated using Gaussian~16. In addition, there are systematic errors resulting from the level of theory and the spatially-localized basis set. 

Since the ionization energy measurements may be affected by the local density of the atom at the center of the sphericalization, the total density was also sphericalized about the "center of mass" of the molecule, computed using the nuclear charges of each atom of glycine as the masses. The ionization energy computed from this sphericalized density, displayed in Fig.~\ref{fig:GlycineCOM}, was 13.9 eV (listed in Table \ref{tab:ionization_energy} under c.m.)  For comparison, the numerical average of the ten individual atomic IPs is 11.5 eV.

\bibliography{SphericalDFT}

\end{document}